\documentclass[iop,usenatbib]{emulateapj}

\usepackage[english]{babel}
\usepackage{graphicx}
\usepackage{fleqn}
\usepackage{amssymb}
\usepackage{amsmath}
\usepackage{booktabs}

\usepackage[breaklinks,colorlinks,citecolor=blue]{hyperref}

\usepackage{anyfontsize}
\usepackage{threeparttable}

\usepackage{txfonts} 

\usepackage{color, colortbl}

\renewcommand{\S}{Section}
\newcommand{\F}{Fig.}

\newcommand{\ve}[1]{\boldsymbol{#1}}
\newcommand{\unit}[1]{\hat{\boldsymbol{#1}}}

\newcommand{\msun}{\mathrm{M}_\odot}
\newcommand{\rsun}{\mathrm{R}_\odot}
\newcommand{\lsun}{\mathrm{L}_\odot}
\newcommand{\au}{\,\textsc{au}}

\newcommand{\renc}{R_\mathrm{enc}}
\newcommand{\mper}{M_\mathrm{per}}
\newcommand{\srel}{\sigma_\mathrm{rel}}

\allowdisplaybreaks

\usepackage{etoolbox}
\newtoggle{ApJFigs}
\togglefalse{ApJFigs}

\begin{document}

\title{Hot Jupiters driven by high-eccentricity migration in globular clusters}

\author{Adrian S. Hamers and Scott Tremaine}
\affil{Institute for Advanced Study, School of Natural Sciences, Einstein Drive, Princeton, NJ 08540, USA}
\email{hamers@ias.edu}

\begin{abstract} 
Hot Jupiters (HJs) are short-period giant planets that are observed around $\sim 1\%$ of solar-type field stars. One possible formation scenario for HJs is high-eccentricity (high-$e$) migration, in which the planet forms at much larger radii, is excited to high eccentricity by some mechanism, and migrates to its current orbit due to tidal dissipation occurring near periapsis. We consider high-$e$ migration in dense stellar systems such as the cores of globular clusters (GCs), in which encounters with passing stars can excite planets to the high eccentricities needed to initiate migration. We study this process via Monte Carlo simulations of encounters with a star+planet system including the effects of tidal dissipation, using an efficient regularized restricted three-body code. HJs are produced in our simulations over a significant range of the stellar number density $n_\star$. Assuming the planet is initially on a low-eccentricity orbit with semimajor axis $1\,\au$, for $n_\star \lesssim 10^3\,\mathrm{pc^{-3}}$ the encounter rate is too low to induce orbital migration, whereas for $n_\star \gtrsim 10^6\,\mathrm{pc^{-3}}$ HJ formation is suppressed because the planet is more likely ejected from its host star, tidally disrupted, or transferred to a perturbing star. The fraction of planets that are converted to HJs peaks at $\approx 2\%$ for intermediate number densities of $\approx 4\times 10^4\,\mathrm{pc^{-3}}$. Warm Jupiters, giant planets with periods between 10 and 100 days, are produced in our simulations with an efficiency of up to $\approx 0.5\%$. Our results suggest that HJs can form through high-$e$ migration induced by stellar encounters in the centers of of dense GCs, but not in their outskirts where the densities are lower. 
\end{abstract}


\section{Introduction}
\label{sect:introduction}
Globular clusters (GCs) are among the oldest and densest stellar systems known. Recent findings have challenged our understanding of these systems. Whereas GCs were traditionally thought to have formed in a single starburst, in the last decade multiple stellar populations have been observed in a large fraction of GCs, and the origin of these populations is still actively debated (see, e.g., \citealt{2012A&ARv..20...50G} for a review). Another puzzling find is that searches for planets around stars in GCs have been unsuccessful. In particular, no planets were found in extensive {\it Hubble Space Telescope} observations of 47 Tuc. This failure was originally interpreted to imply that the occurrence rate of short-period planets around stars in 47 Tuc is at least an order of magnitude lower than for field stars \citep{2000ApJ...545L..47G}. However, a recent study by \citet{2017AJ....153..187M} has shown that this estimate should be revised, given the now better-known radius and period relations of giant planets around field stars; they find that the expected number of planets is only $2.2^{+1.6}_{-1.1}$, so the null result is marginally consistent with the abundance of planets around field stars \citep{2017AJ....153..187M}. Similar arguments apply to other surveys of planets in GCs, in particular the ground-based surveys of 47 Tuc by \citet{2005ApJ...620.1043W} and of $\omega$ Cen by \citet{2008ApJ...674.1117W}. Therefore, the existence of short-period giant planets in GCs is still an open question.

A deficit of short-period planets in GCs would suggest that planet formation in GCs is inefficient, and/or that dynamical interactions in these dense environments destroy such planets after they form. Inhibited planet formation might, for example, be due to radiation from nearby massive stars \citep{2000A&A...362..968A,2004ApJ...611..360A,2013MNRAS.431...63T}. Alternatively, GCs have low metallicities and the giant-planet occurrence rate is known to correlate with metallicity in field stars \citep{2001A&A...373.1019S,2005ApJ...622.1102F}. However, it is unclear whether this relation also applies to GCs, whose formation history is not well-understood.

Dynamical interactions can disrupt planetary systems (e.g., \citealt{1992ApJ...399L..95S}), but they can also enhance the numbers of short-period planets by exciting high planetary eccentricities: in particular, if the periapsis of a giant-planet orbit becomes as small as a few stellar radii, tidal dissipation in the planet excited by interactions with the host star may become strong enough to drive migration of the planet to a tight circular orbit with a period of a few days, creating a hot Jupiter (HJ) --- the easiest class of planet to detect in transit surveys. Once the planet has migrated to such a tightly bound orbit, it becomes immune to further perturbations from passing stars\footnote{In some cases, further orbital decay driven by tidal dissipation in the star could shrink the orbit until the planet is disrupted by the star. However, the relatively high occurrence rate of HJs around field stars indicates that stellar tidal dissipation is typically inefficient in solar-type stars, and we shall ignore this process in the present paper.}.

High-eccentricity (high-$e$) migration around field stars has been widely studied. The possible eccentricity excitation mechanisms include planet-planet scattering \citep{1996Sci...274..954R,2008ApJ...686..580C,2008ApJ...686..621F,2008ApJ...686..603J,2008ApJ...678..498N,2012ApJ...751..119B}; Lidov-Kozai (LK) oscillations  \citep{1962P&SS....9..719L,1962AJ.....67..591K} in binary-star systems  \citep{2003ApJ...589..605W,2007ApJ...669.1298F,2012ApJ...754L..36N,2015ApJ...799...27P,2016MNRAS.456.3671A,2016ApJ...829..132P,2017ApJ...835L..24H}, triple-star systems \citep{2017MNRAS.466.4107H,2017arXiv171005920G}, and multiplanet systems \citep{2015ApJ...805...75P,2016ApJ...820...55X}; and secular chaos in multiplanet systems \citep{2011ApJ...735..109W,2011ApJ...739...31L,2014PNAS..11112610L,2017MNRAS.464..688H}. Although many authors have studied the dynamics of planets in open clusters and GCs (e.g., \citealt{1992ApJ...399L..95S,1997A&A...326L..21D,2001MNRAS.322..859B,2001MNRAS.324..612D,2006ApJ...640.1086F,2009ApJ...697..458S,2011MNRAS.411..859M,2012ApJ...754...57B,2012MNRAS.427.1587C,2012MNRAS.419.2448P,2013MNRAS.433..867H,2013ApJ...772..142L,2015MNRAS.448..344L,2015MNRAS.449.3543W,2015MNRAS.453.2759Z,doi:10.1093/mnras/stx1464}), the interplay between perturbations from passing stars and dissipative planetary tides has scarcely been investigated. \citet{2016ApJ...816...59S} considered the formation of HJs in two-planet systems in open clusters and found that HJ formation occurs in $\sim 1\%$ of the planetary systems. To our knowledge, no study has focused on similar processes in the much denser GCs.

In this paper, we study the formation of HJs in GCs through high-$e$ migration induced by passing stars. A computational challenge in this problem is the wide range of timescales: from a few days for the orbital period of an HJ, to 30 yr for the encounter time of a star with impact parameter $30\au$ and relative velocity $5\,\mathrm{km\,s^{-1}}$, to $10^{10}\,\mathrm{yr}$ for the lifetime of a GC. We approach this problem by using an efficient regularized restricted three-body code that includes tides and general relativistic corrections. This method allows us to simulate the cumulative effect of encounters over the lifetime of a GC, which is a prohibitive endeavor using general-purpose direct $N$-body integrators. 

The plan of the paper is as follows. In \S\,\ref{sect:meth}, we present our regularized restricted three-body code and test it using three-body integrations with a highly accurate $N$-body code. In \S\,\ref{sect:pop_syn}, we apply our method to a population-synthesis study of planets in the centers of dense GCs, and we describe the properties of the migrating planets. In addition, we present an analytic model that approximately describes the period distribution of the HJs that are formed in the simulations. We discuss our results in \S\,\ref{sect:discussion}, and we conclude in \S\,\ref{sect:conclusions}.

\section{Methodology}
\label{sect:meth}

\subsection{Numerical integration method}
\label{sect:meth:des}

Consider a star of mass $M_\star$ orbited by a planet with semimajor axis $a$ and mass $M_\mathrm{p}$. The orbit of the planet, also referred to as the ``binary'', is perturbed by a passing star with mass $\mper$. We assume that $M_\mathrm{p} \ll M_\star,\mper$, in which case the planet can be interpreted as a test particle that does not affect the stellar motion. The effect of the encounter on the planetary orbit can be calculated numerically using direct $N$-body integration. However, rather than using standard integration methods, we take advantage of the fact that we are dealing with a restricted three-body problem. In particular, we assume that the perturber moves on a hyperbolic orbit having separation $\ve{R}(t)$ with respect to the planet-hosting star, and we regularize the motion of the planet with respect to its host star. We use Kustaanheimo-Stiefel (KS) regularization \citep{ks65,1971lrcm.book.....S} with the time transformation $\mathrm{d} t/\mathrm{d} s = r$, where $t$ and $s$ are the physical and fictitious times, respectively, and $r$ is the planet-host star separation. This approach allows us to compute the effect of the perturber on the planetary orbit with a factor of $\sim$ 20-100 performance increase compared to a direct $N$-body code, without much loss of accuracy (see \S\,\ref{sect:meth:val} below).

In addition to the gravitational force from the point-mass perturber, we also include general relativistic corrections to the first post-Newtonian (PN) order and tidal effects induced in the planet by its host star. The latter are required to be able to model the formation of HJs. The perturbing acceleration $\ve{P}$ to the regularized motion of the planet is given by
\begin{align}
\label{eq:P}
\ve{P} = \ve{P}_\mathrm{per} + \ve{P}_\mathrm{tides} + \ve{P}_\mathrm{1PN}.
\end{align}
Here, $\ve{P}_\mathrm{per}$ is given by
\begin{align}
\label{eq:P_per}
\ve{P}_\mathrm{per} = -G \mper \left ( \frac{ \ve{r} - \ve{R}}{||\ve{r}-\ve{R}||^3} + \frac{\ve{R}}{||\ve{R}||^3} \right ),
\end{align}
where $\ve{r}$ and $\ve{R}$ are, respectively, the relative separation vectors of the planet and the perturber with respect to the host star; the first and second terms in equation~(\ref{eq:P_per}) are the `direct' and `indirect' terms, respectively. 

We take into account the planetary tidal evolution with the equilibrium tide model. Here, we assume that the planetary rotational spin vector is always aligned with the normal to the planetary orbit. Generally, the expression for the tidal perturbing acceleration is then (cf.\ Eq.\ 8 of \citealt{1981A&A....99..126H})
\begin{align}
\label{eq:P_tides}
\ve{P}_\mathrm{tides} = - 3 k_\mathrm{AM,p} \frac{ G M_\star^2}{M_\mathrm{p} r^2} \left ( \frac{R_\mathrm{p}}{r} \right )^5 \left [ \left ( 1 + 3 \tau_\mathrm{p} \frac{\dot{r}}{r} \right ) \unit{r} - \left ( \Omega_\mathrm{p} - \dot{\theta} \right ) \tau_\mathrm{p} \unit{\theta} \right ],
\end{align}
where $R_\mathrm{p}$, $k_\mathrm{AM,p}$, and $\tau_\mathrm{p}$ are the planetary radius, apsidal motion constant, and tidal time lag; the overhead dot denotes time derivatives, $\unit{r}\equiv \ve{r}/r$, $\theta$ is the true anomaly, $\unit{\theta}$ is the azimuthal unit vector in the orbital plane, pointing in the direction of increasing $\theta$, and $\Omega_\mathrm{p}$ is the planetary spin frequency. The planet's rotational angular momentum is much smaller than its orbital angular momentum; therefore, the tidal torque is small as well. This justifies the assumption that the orbit-averaged tidal torque is zero, i.e., $\langle \ve{r} \times \ve{P}_\mathrm{tides} \rangle = \ve{0}$, which is equivalent to $\left \langle (\Omega_\mathrm{p} - \dot{\theta})/r^6 \right \rangle = 0$ or $\langle \Omega_\mathrm{p} \rangle = \Omega_\mathrm{PS}(e)$, where $\Omega_\mathrm{PS}(e)$ is a function of eccentricity given by
\begin{align}
\label{eq:Omega_PS}
\Omega_\mathrm{PS}(e) = n \frac{1 + \frac{15}{2} e^2 + \frac{45}{8} e^4 + \frac{5}{16} e^6}{ \left ( 1+3e^2+\frac{3}{8} e^4\right ) \left (1-e^2 \right )^{3/2}},
\end{align}
with $n=\sqrt{G M_\star/a^3}$ the mean motion. Equation~(\ref{eq:Omega_PS}) is equivalent to pseudosynchronization, i.e., $\langle \dot{\Omega}_\mathrm{p} \rangle=0$ (cf. equation 42 of \citealt{1981A&A....99..126H}). In the numerical integrations, we set $\Omega_\mathrm{p} = \Omega_\mathrm{PS}(e)$, where $e$ is the instantaneous eccentricity. We note that this treatment of tides is slightly different from \citeauthor{2011ApJ...735..109W} (\citeyear{2011ApJ...735..109W}; cf. Eq. A7) and \citeauthor{2016AJ....152..174A} (\citeyear{2016AJ....152..174A}; cf. Eq. 6), who set the term in $\ve{P}_\mathrm{tides}$ proportional to $\unit{\theta}$ to zero. 

Lastly, the relativistic perturbing acceleration is \citep{1938AnMat..39...65E}
\begin{align}
\ve{P}_\mathrm{1PN} = \frac{G M_\star}{c^2 r^3} \left [4 \left ( \ve{r} \cdot \dot{\ve{r}} \right ) \dot{\ve{r}} +4\, G M_\star \, \unit{r} - \left( \dot{\ve{r}} \cdot \dot{\ve{r}} \right ) \ve{r} \right ].
\end{align}

\subsection{Validation}
\label{sect:meth:val}

\begin{figure*}
\center
\iftoggle{ApJFigs}{
\includegraphics[scale = 0.43, trim = 45mm -5mm 0mm 0mm]{elements_Q_1.0.eps}
\includegraphics[scale = 0.43, trim = 45mm -5mm 0mm 0mm]{elements_Q_3.0.eps}
\includegraphics[scale = 0.43, trim = 45mm -5mm 0mm 0mm]{elements_Q_10.0.eps}
}{
\includegraphics[scale = 0.43, trim = 45mm -5mm 0mm 0mm]{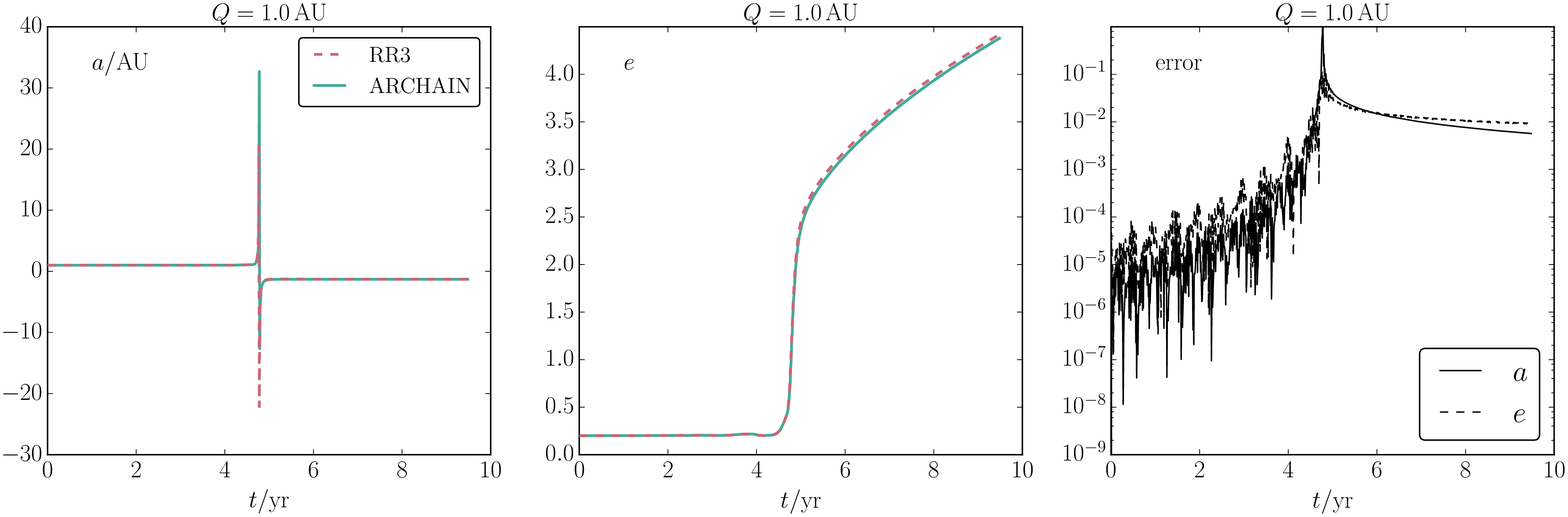}
\includegraphics[scale = 0.43, trim = 45mm -5mm 0mm 0mm]{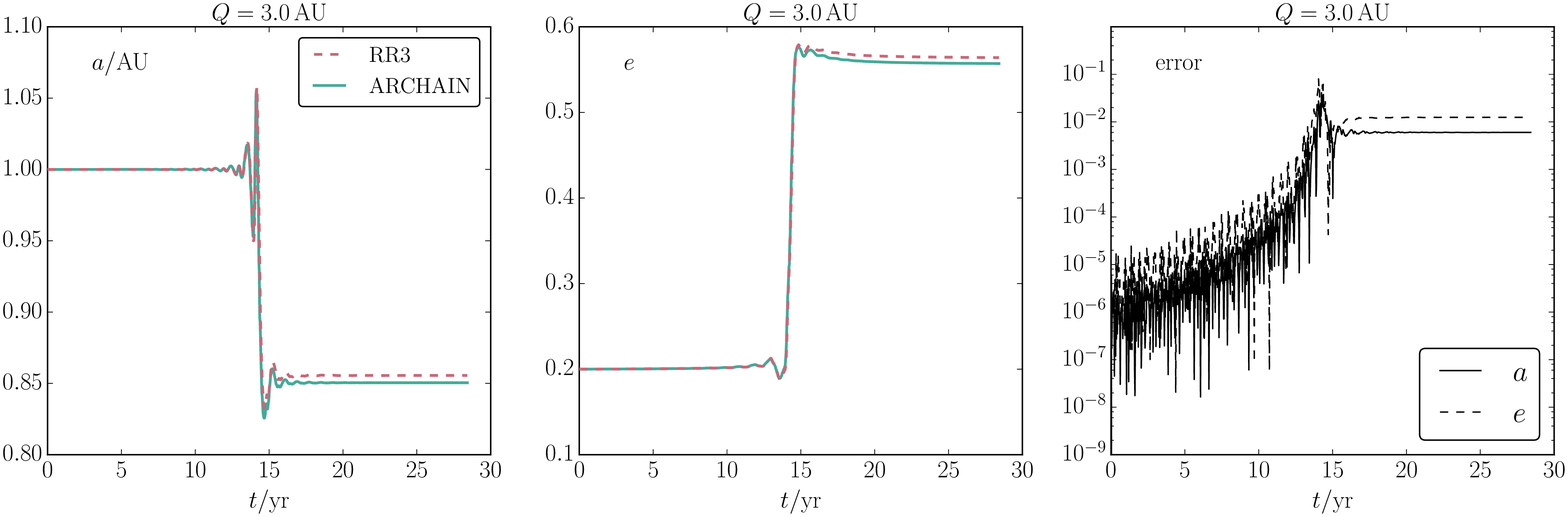}
\includegraphics[scale = 0.43, trim = 45mm -5mm 0mm 0mm]{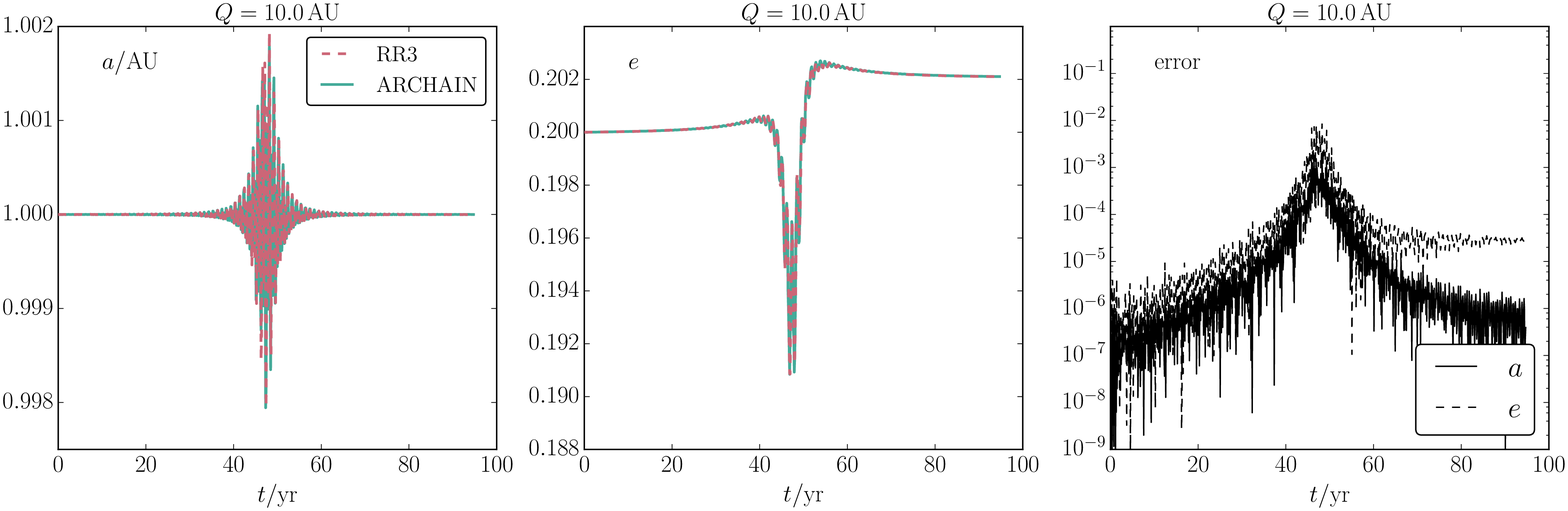}
}
\caption {Comparison of the regularized restricted 3-body code (\textsc{RR3}; red dashed lines) with 3-body integrations carried out with \textsc{ARCHAIN} (green solid lines). The left and middle panels show the planetary semimajor axis and eccentricity, respectively, as a function of time. The right panels show the absolute value of the relative error in the semimajor axis (black solid lines) and the eccentricity (black dashed lines) in \textsc{RR3} relative to \textsc{ARCHAIN}. Each row corresponds to a different value of the perturber's closest approach $Q$, indicated above each panel. }
\label{fig:val}
\end{figure*}

We test our regularized restricted three-body code (\textsc{RR3}) with integrations carried out with \textsc{ARCHAIN} \citep{2006MNRAS.372..219M,2008AJ....135.2398M}, a high-accuracy $N$-body code that uses chain regularization. We use \textsc{ARCHAIN} implemented within \textsc{AMUSE} \citep{2013CoPhC.183..456P,2013A&A...557A..84P}. In all validation integrations shown in this Section, we only include Newtonian point-mass dynamics, i.e., we set $\ve{P}_\mathrm{tides} = \ve{P}_\mathrm{1PN} = \ve{0}$ in equation~(\ref{eq:P}), and the speed of light in \textsc{ARCHAIN} is set to $10^{100}\,c$ (i.e., effectively infinity) to eliminate relativistic effects. In contrast, in the simulations of \S\,\ref{sect:pop_syn} below, both tidal and general relativistic effects are taken into account. 

The following initial conditions are assumed, representing some of the strong and weak encounters in the population-synthesis calculations of \S\,\ref{sect:pop_syn}. The masses are set to $M_\star = 1\, \msun$, $M_\mathrm{p} = 1 \, M_\mathrm{J}$ (the Jupiter mass $M_\mathrm{J}\simeq0.001\,\msun$), and $\mper = 1\,\msun$. The initial binary semimajor axis and eccentricity are $a = 1\,\au$ and $e=0.2$. The perturber's velocity at infinity is $V_\infty = 10\, \mathrm{km\,s^{-1}}$, and its periapsis distance $Q$ is taken to be either 1, 3, or 10 $\au$. The binary's inclination with respect to the perturber is zero (i.e., the binary orbit is prograde relative to the perturber), and the longitudes of periapsis of the binary and perturber initially differ by $90^\circ$. The initial true anomaly of the binary is zero. The duration of the integrations is $\Delta t = 20 \, Q/V_\infty$ with the periapsis passage occurring at $t=10 \, Q/V_\infty$; for the adopted values of $Q$, $Q/V_\infty \simeq 0.47$, 1.4 and 4.7 yr, respectively. Note that these tests are particularly challenging because (i) the orbits are coplanar and (ii) when $Q=1\au$ the unperturbed orbits of the planet and the perturber can collide (in the validation tests, we assume that the radii of all three bodies are zero and hence do not check for the occurrence of collisions). 

In the left and middle panels of \F\,\ref{fig:val}, we show the semimajor axis and eccentricity as a function of time. The right panels show the relative errors in the semimajor axis and the eccentricity in \textsc{RR3} as a function of time, assuming the \textsc{ARCHAIN} code gives the exact result. Each row corresponds to a different value of $Q$, indicated above each panel. 

The encounter with $Q=1\,\au$ results in a destructive perturbation to the planetary orbit, i.e., the planet becomes unbound from its host star ($a<0$ and $e\geq1$). Despite the large perturbation, \textsc{RR3} computes the final semimajor axis and eccentricity with $\sim 1\%$ error compared to \textsc{ARCHAIN}. The performance increase with respect to \textsc{ARCHAIN} is a factor of $\sim 20$. 

The encounter with $Q=3\,\au$ is less destructive, leaving the planet bound to its host star. The eccentricity increases from 0.2 to $\approx 0.55$ and the semimajor axis decreases by $\approx 15\%$. The error with respect to \textsc{ARCHAIN} is again $\sim 1\%$, and the performance increase is a factor of $\sim 100$. 

Lastly, the case $Q=10\,\au$ corresponds to a `secular' encounter in which the angular speed of the perturber at periapsis is much lower than the binary mean motion, i.e., the associated ratio of these quantities,
\begin{align}
\label{eq:R_def}
\mathcal{R} &= \left [ \left ( 1 + \frac{\mper}{M_\star} \right ) \left (\frac{a}{Q} \right )^3 \left ( 2 + \frac{Q V_\infty^2}{G(\mper + M_\star)} \right ) \right ]^{1/2}
\end{align}
satisfies $\mathcal{R} \ll 1$ (in this formula, we assumed $M_\mathrm{p} \ll M_\star$). For $Q=10\,\au$, $\mathcal{R} \simeq 0.072$, indicating that the encounter is of the secular type (in contrast, $\mathcal{R} \simeq 2.0$ and $\simeq0.40$ for $Q=1$ and 3 $\au$, respectively). Secular encounters produce a permanent change in the eccentricity but no permanent change in semimajor axis. This is indeed the case in the bottom row of \F\,\ref{fig:val}: the semimajor axis returns to its original value after being perturbed by $\sim0.2\%$, whereas the eccentricity is changed by $\approx 2\%$ after the encounter. The error made with respect to \textsc{ARCHAIN} is less than $0.1\%$, whereas the performance gain is a factor of $\sim 100$. Generally, the performance gain with \textsc{RR3} tends to increase for more secular encounters.

\section{Population-synthesis study}
\label{sect:pop_syn}

\subsection{Gravitational dynamics and tidal evolution}
\label{sect:pop_syn:dyn}

We apply the regularized restricted three-body code \textsc{RR3} described in \S\,\ref{sect:meth} to model numerically the effects of encounters on planetary orbits in dense stellar systems such as GCs. The planet is assumed to have an initial semimajor axis $a_0=1$, 2, or 4 $\au$. These values approximately span the range in which most known giant exoplanets are found, although many as-yet undetected giant planets are likely to be present at larger semimajor axes. The initial planetary eccentricity $e$ is assumed to follow from a Rayleigh distribution with an rms value of 0.33 \citep{2008ApJ...686..603J}, cut off at a maximum eccentricity of 0.6. The resulting distribution has an rms value of $\simeq 0.31$, which is close to the rms value of $\simeq 0.32$ for planets with periods above 10 days around stars in the solar neighborhood\footnote{Rms value obtained from \href{www.exoplanets.org}{\it www.exoplanets.org} on October 30 2017.}. We  checked that there is little to no dependence of our results on the initial eccentricity.

In addition to the Newtonian accelerations due to the host and perturbing stars, we include the lowest-order precession of the planetary apsides due to general relativity and tidal evolution of the planetary orbit. These effects are implemented with the assumptions described in \S\,\ref{sect:meth:des}, and the associated parameters $R_\mathrm{p}$, $\tau_\mathrm{p}$, and $k_\mathrm{AM,p}$ are given in the top part of Table \ref{table:par}. The tidal time lag $\tau_\mathrm{p}$ is assumed to be constant \citep{2012arXiv1209.5723S}. We set $\tau_\mathrm{p} = 0.66 \, \mathrm{s}$, for which an HJ with a 5-day orbital period is circularized in less than 10 Gyr \citep{2012arXiv1209.5724S}. This time lag corresponds to a tidal quality factor $Q_\mathrm{p}\simeq 1.1\times 10^5$ (cf.\ equation 37 from \citealt{2012arXiv1209.5724S}), and $Q_\mathrm{p} \propto 1/\tau_\mathrm{p}$. We assume that stellar tides are negligible.

\subsection{Generating encounters}
\label{sect:pop_syn:enc_gen}
\subsubsection{Flux of perturbing stars}
\label{sect:pop_syn:enc_gen:flux}

In our simulations, encounters are generated continuously until the current age of the GC is reached (assumed to be 10 Gyr). To sample the encounters, we assume a locally homogeneous stellar background with stellar number density $n_\star$ and one-dimensional velocity dispersion $\sigma$ independent of stellar mass. We assume a Maxwellian stellar velocity distribution at large distances from the host star, such that the distribution function is
\begin{align}
\label{eq:DF1}
\mathrm{DF} \propto n_\star f(\mper) \exp \left ( -\frac{v^2}{2\srel^2} \right ),
\end{align}
where $f(\mper)\,\mathrm{d}\mper$ is the fraction of stars with masses in the interval $\mathrm{d} \mper$ and $\srel= \sqrt{2} \, \sigma$ is the relative velocity dispersion \citep{2008gady.book.....B}.  When the gravitational attraction of the host star is included, the distribution function must still be a function of the energy $E = \frac{1}{2} v^2 - G M /r$, where $M\equiv M_\star + \mper$ and $r$ is the distance from the host star. Therefore, the distribution function is modified to
\begin{align}
\label{eq:DF2}
\mathrm{DF} &\propto n_\star f(\mper) \exp \left ( -\frac{v^2}{2\srel^2} \right ) \exp \left ( \frac{ GM}{r \srel^2} \right )
\end{align}
for $v > \sqrt{2G(M_\star+\mper)/r}$, and zero otherwise. 

We now introduce an imaginary `encounter sphere' centered at the host star with a radius $\renc \gg a$. We also choose $\renc \ll R_\mathrm{const}$, where $R_\mathrm{const}$ is the length scale beyond which $n_\star$ and $\sigma$ are no longer (approximately) constant. Stars impinging on the encounter sphere are considered to be perturbers. Using equation~(\ref{eq:DF2}), we find that the number density of perturbers at the encounter sphere within a mass range $\mathrm{d} \mper$ and with velocities between $(v_x,v_y,v_z)$ and $(v_x+\mathrm{d}v_x,v_y+\mathrm{d}v_y,v_z+\mathrm{d}v_z)$ is 
\begin{align}
\label{eq:dn}
\nonumber &\mathrm{d} {n_{\star,\mathrm{enc}}} = \frac{ n_\star}{(2\pi \srel^2 )^{3/2}} f(\mper) \mathrm{d} \mper \exp \left ( \frac{GM}{\renc \srel^2} \right ) \\
&\quad \times H\left(v^2-\frac{2GM}{\renc} \right ) \exp \left ( -\frac{v^2}{2 \srel^2} \right ) \, \mathrm{d} v_x \, \mathrm{d} v_y \, \mathrm{d} v_z,
\end{align}
where $H(x)$ is the Heaviside step function. Integration of equation~(\ref{eq:dn}) over all perturber masses and velocities gives
\begin{align}
\label{eq:n}
n_{\star,\mathrm{enc}} = n_\star \int \, \mathrm{d} \mper f(\mper) \, W \left [ \frac{G(M_\star + \mper)}{\renc \srel^2} \right ],
\end{align}
where
\begin{align}
\label{eq:W}
W(x) \equiv 2\sqrt{x/\pi} + \exp(x) \, \mathrm{erfc}(\sqrt{x}),
\end{align}
and $\mathrm{erfc}(x) = 1 - \mathrm{erf}(x)$ is the complementary error function. Therefore, the fraction of perturbers at the encounter sphere with mass $\mper$ is proportional to $W(x)f(\mper)$; we will use this result in our sampling procedure described below (\S\,\ref{sect:pop_syn:enc_gen:sampling}). Equation~(\ref{eq:n}) shows that the stellar number density at the encounter sphere, $n_{\star,\mathrm{enc}}$, is larger than $n_\star$ due to gravitational focusing (note that $W(x) \geq 1$ for $x \geq 0$). Henceforth,  when we use the term ``number density'' we will always be referring to $n_\star$.

Consider a point on the encounter sphere with position vector $\ve{R}_\mathrm{enc}$ relative to the host star. Next, define a local coordinate system centered on this point in which the $z$ axis is directed toward the host star, i.e., $\hat{\boldsymbol{z}} = - \unit{R}_\mathrm{enc}$, and the $x$ and $y$ axes lie on the tangent plane of $\ve{R}_\mathrm{enc}$ on the encounter sphere. The differential flux of stars into the encounter sphere is given by $\mathrm{d} F = v_z H(v_z) \,\mathrm{d} n_{\star,\mathrm{enc}}$ \citep{1972A&A....19..488H}, i.e.,
\begin{align}
\label{eq:vel_distr}
\nonumber &\mathrm{d} F = \frac{ n_\star}{(2\pi \srel^2 )^{3/2}} f(\mper) \mathrm{d} \mper \exp \left ( \frac{GM}{\renc \srel^2} \right ) H(v_z) v_z \\
&\quad \times H\left(v^2-\frac{2GM}{\renc} \right ) \exp \left ( -\frac{v^2}{2 \srel^2} \right ) \, \mathrm{d} v_x \, \mathrm{d} v_y \, \mathrm{d} v_z.
\end{align}
Integrating the differential flux over all perturber masses, velocities, and the entire encounter sphere, we obtain a total encounter rate of
\begin{align}
\label{eq:Gamma}
\nonumber \Gamma &= 2 \sqrt{2\pi} \renc^2 n_{\star} \srel \\
&\quad \times \int \mathrm{d} \mper f(\mper ) \left [1 + \frac{G(M_\star+\mper)}{\renc \srel^2} \right ].
\end{align}
In the limit of large $\renc$ ($\renc \gg GM/\srel^2$, i.e., weak or negligible gravitational focusing), equation~(\ref{eq:Gamma}) reduces to
\begin{align}
\label{eq:Gamma_limit}
\nonumber \Gamma &\approx 2 \sqrt{2\pi} \renc^2 n_{\star} \srel \int \mathrm{d} \mper f(\mper ) \\
&= 2 \sqrt{2\pi} \renc^2 n_{\star} \srel,
\end{align}
independent of the perturber mass function. 

\subsubsection{Numerical sampling procedure}
\label{sect:pop_syn:enc_gen:sampling}

We generate encounters using the following procedure. 

\begin{enumerate}

\item The perturber mass $\mper$ is assumed to follow a Salpeter distribution \citep{1955ApJ...121..161S}, modified to account for the finite lifetime of stars and the gravitational focusing implied by equation~(\ref{eq:n}). 

Specifically, an initial mass $M_\mathrm{i}$ is sampled from a Salpeter distribution, $\mathrm{d}N/\mathrm{d}M_\mathrm{i}\propto M_\mathrm{i}^{-2.35}$, with lower and upper limits 0.1 and 100 $\msun$. Using the \textsc{SSE} stellar evolution code \citep{2000MNRAS.315..543H} as implemented in \textsc{AMUSE} \citep{2013CoPhC.183..456P,2013A&A...557A..84P} and assuming a metallicity $Z=0.001$, this initial mass is replaced by the mass after 5 Gyr of stellar evolution, $M_\mathrm{f}$. Subsequently, we compute the associated value of $x = G(M_\star+M_\mathrm{f})/(\renc \srel^2)$ and $W(x)$ (eq.~\ref{eq:W}), and reject the sampled mass if $W(x)/W_\mathrm{max} < y$, where $y$ is a random number between 0 and 1, and $W_\mathrm{max}$ is the maximum value of $W$ over the allowed range of $M_\mathrm{f}$. 

We do not account for the possibility that stars may be ejected from the cluster due to asymmetric mass loss or other effects, nor do we account for binaries among the perturbers. 

\item A random location $\ve{R}_\mathrm{enc}$ on the encounter sphere is chosen.  The velocities $v_x$, $v_y$, and $v_z$ are then sampled from the distribution implied in equation~(\ref{eq:vel_distr}). From these velocities, the periapsis distance $Q$ and the speed at infinity, $V_\infty$, are computed. From the velocities and $\ve{R}_\mathrm{enc}$ the orientation of the hyperbolic orbit is determined.  

\item The next encounter is generated assuming that the probability for that the time delay between encounters exceeding $\Delta t$ is $\mathrm{exp}(- \Gamma \Delta t)$, with $\Gamma$ given by equation~(\ref{eq:Gamma}).  

\end{enumerate}

\subsubsection{Planetary perturbations and the encounter sphere radius}
\label{sect:pop_syn:enc_gen:per}

The gravitational effect of each encounter on the planetary orbit is followed from the time the perturber enters the encounter sphere until the time that it again impinges on the encounter sphere on the opposite side of the hyperbolic orbit. The effects of the perturber on the planet when the perturber is outside the encounter sphere are neglected. Note that we do take into account the attraction of the host star on the perturber at all distances by including the effects of gravitational focusing as described in \S\,\ref{sect:pop_syn:enc_gen:flux}.

In principle, the radius of the encounter sphere $\renc$ could be taken to be close to $R_\mathrm{const}$ such that the gravitational effect of each encounter on the planetary orbit is fully accounted for. However, this approach is computationally impractical. Fortunately, gravitational forces from perturbers at large distances contribute negligibly to orbital changes because tidal forces fall off as $1/R^3$ (see equation~\ref{eq:P_per}).  Therefore, it is sufficient to restrict $\renc$ to relatively small values. In the simulations below, we vary $\renc$ between 25 and 100 $\au$ and show that these encounter radii are large enough for our purposes (e.g., the outcome fractions are largely independent of $\renc$).

We assume that there is at most one perturber within the encounter sphere at a given time. This is justified because in our simulations, the typical timescale for a perturber to pass through the encounter sphere is short compared to the timescale for the next perturber to enter the encounter sphere. More quantitatively, let the encounter passage time be estimated by\footnote{This analysis assumes that the trajectory of the perturber relative to the binary is not strongly perturbed by the encounter, that is $b\gg G(\mper+M_\star)/V_\infty^2$ where $b$ is the impact parameter. This assumption is not correct for the closest encounters but should be adequate for this argument.} $\Delta t_\mathrm{passage} \simeq \renc/\srel$, and the time to the next encounter by $\Delta t_\mathrm{enc} \sim 1/\Gamma$, with $\Gamma$ estimated from equation~(\ref{eq:Gamma_limit}). Then,
\begin{align}
\label{eq:ratio_t_pas_t_enc}
\nonumber \frac{\Delta t_\mathrm{passage}}{\Delta t_\mathrm{enc}} &\sim 2\sqrt{2\pi} \, \renc^3 n_\star \simeq 5.7 \times 10^{-4} \, \left ( \frac{\renc}{100\,\au} \right )^3 \\
&\quad \times \left ( \frac{n_\star}{10^6\,\mathrm{pc^{-3}}} \right ),
\end{align}
where we substituted numerical values corresponding to the largest ratio $\Delta t_\mathrm{passage}/\Delta t_\mathrm{enc}$ in the simulations. This shows that our assumption of at most one perturber in the encounter sphere is justified for our simulations. We also assume that the orbital phase of the planet is randomized each time a new perturber is introduced in the encounter sphere, which is justified by a similar argument:
\begin{align}
\nonumber \frac{P_\mathrm{orb}}{\Delta t_\mathrm{enc}} &\sim 2 \sqrt{ \frac{(2\pi a)^3}{G M_\star}} \renc^2 n_\star \srel \\
\nonumber &\simeq 1.0 \times 10^{-5} \, \left ( \frac{a}{1\,\au} \right )^{\frac{3}{2}} \left ( \frac{M_\star}{1\,\msun} \right )^{-\frac{1}{2}} \left ( \frac{\renc}{100\,\au} \right )^2 \\
&\quad \times \left ( \frac{n_\star}{10^6\,\mathrm{pc^{-3}}} \right ) \left ( \frac{\srel}{\sqrt{2} \times 6 \, \mathrm{km\,s^{-1}}} \right ),
\end{align}
where $P_\mathrm{orb}$ is the orbital period of the planet.

\subsubsection{Isolated tidal evolution}
\label{sect:pop_syn:enc_gen:tides}
We assume that the orbit of the planet evolves due to tides only when there are no perturbing stars within the encounter sphere. In this process, the semimajor axis and eccentricity of the planet are evolved according to the orbit-averaged version of equation~(\ref{eq:P_tides}),
\begin{subequations}
\label{eq:TF_av}
\begin{align}
\label{eq:TF_av_a}
\frac{\mathrm{d} a}{\mathrm{d} t} &= - 21 \, k_\mathrm{AM,p} n^2 \tau_\mathrm{p} \frac{M_\star}{M_\mathrm{p}} \left ( \frac{R_\mathrm{p}}{a} \right )^5 a e^2 \frac{f(e)}{\left(1-e^2\right)^{15/2}}; \\
\label{eq:TF_av_e}
\frac{\mathrm{d} e}{\mathrm{d} t} &= -\frac{21}{2} k_\mathrm{AM,p} n^2 \tau_\mathrm{p} \frac{M_\star}{M_\mathrm{p}} \left ( \frac{R_\mathrm{p}}{a} \right )^5 e \frac{f(e)}{\left(1-e^2\right)^{13/2}}, 
\end{align}
\end{subequations}
where
\begin{align}
f(e) = \frac{1 + \frac{45}{14} e^2 + 8 e^4 + \frac{685}{224} e^6 + \frac{255}{448} e^8 + \frac{25}{1792} e^{10}}{1 + 3 e^2 + \frac{3}{8} e^4}.
\end{align}
Note that equations~(\ref{eq:TF_av_a}) and (\ref{eq:TF_av_e}) conserve the semilatus rectum, $a\left(1-e^2\right)$, and consequently, the orbital angular momentum. Also, note that equations~(\ref{eq:TF_av_a}) and (\ref{eq:TF_av_e}) can be obtained from equations (9) and (10), respectively of \citet{1981A&A....99..126H}, by replacing $\Omega$ in the latter equations by its pseudosynchronous value, equation~(\ref{eq:Omega_PS}).

\begin{table}
\begin{threeparttable}
\begin{tabular}{lp{3.2cm}c}
\toprule
Symbol & Description & (Range of) Value(s) \\
\toprule
\multicolumn{3}{l}{Planetary System} \\
\midrule
$M_\star$					& Mass of planet-hosting star					& $1 \, \msun$ \\
$M_\mathrm{p}$			& Planetary mass							& $1 \, M_\mathrm{J}\simeq9.55\times10^{-4}\,\msun$ \\
$R_\mathrm{p}$			& Planetary radius							& $1 \, R_\mathrm{J}\simeq7.15\times10^9\,\mathrm{cm}$ \\
$\tau_\mathrm{p}$        		& Planetary tidal time lag		                    		& $0.66 \, \mathrm{s}$ \\
$k_\mathrm{AM,p}$           	& Planetary apsidal motion constant                    	& 0.25 \\
$\theta_\star$          			& Initial stellar obliquity (stellar spin-planetary orbit angle)  		& $0^\circ$ \\
$a_0$                       			& Initial planetary orbital semimajor axis 			& 1, 2, 4 $\au$ \\
$e_0$                       			& Initial planetary orbital eccentricity 				& 0.01--0.6 ${}^\mathrm{a}$ \\
\toprule
\multicolumn{3}{l}{Cluster and Encounter Properties} \\
\midrule
$\mper$                   	& Perturber mass                  					& 0.1-100 $\mathrm{M}_\odot$ ${}^\mathrm{b}$ \\
$n_\star$ 					& Stellar number density						& $10^{3-6} \, \mathrm{pc^{-3}} $ ${}^\mathrm{c}$ \\
$\sigma$ 					& Stellar (not relative) velocity dispersion 			& $6\,\mathrm{km \, s^{-1}}$ \\
$\tau_\mathrm{age}$ 		& Cluster age 								& $10 \,\mathrm{Gyr}$ \\
$z$						& Cluster metallicity							& 0.001 \\
$\renc$ 			& Encounter sphere radius					& 25, 50, 75, 100 $\au$ \\
\midrule
\multicolumn{3}{l}{Derived Encounter Quantities} \\
\midrule
$\rho_\star$ 				& Stellar mass density 						& $2.6 \times 10^{2-5} \, \msun \mathrm{pc^{-3}}$ ${}^\mathrm{d}$ \\
$l_\star$					& Stellar luminosity density					& $4.5 \times 10^{2-5} \, \lsun \mathrm{pc^{-3}}$ ${}^\mathrm{e}$ \\
$Q_\mathrm{min,est}$		& Estimate of the smallest periapsis distance		& (0.0068 -- 5.1) $ \au$ ${}^\mathrm{f}$ \\
$N_\mathrm{enc,est}$ 		& Estimate of the total number of encounters 		& 10 -- $1.2\times 10^5$ ${}^\mathrm{g}$\\
\bottomrule
\end{tabular}
\begin{tablenotes}
            \item[a] Rayleigh distribution with rms 0.33; subsequently cut off at 0.6.
            \item[b] Salpeter initial mass function \citep{1955ApJ...121..161S}, corrected for a stellar age of 5 Gyr (metallicity $z=0.001$), and for gravitational focusing (see \S\,\ref{sect:pop_syn:enc_gen}).
            \item[c] 20 values with logarithmic spacing.
            \item[d] Computed from the final mass function assumed for the perturbers.
            \item[e] Computed from the final mass and luminosity functions assumed for the perturbers.
            \item[f] Minimum and maximum values for the assumed parameters; computed using equation~(\ref{eq:Q_min_est}).
            \item[g] Minimum and maximum values for the assumed parameters; computed according to $N_\mathrm{enc,est}=\Gamma \tau_\mathrm{age}$ with $\Gamma$ given by equation~(\ref{eq:Gamma}).
\end{tablenotes}
\caption{Description of symbols used and assumed values in the simulations. Top part: properties of the planet and planet-hosting star. Middle and bottom parts: cluster and encounter properties, and some derived quantities. }
\label{table:par}
\end{threeparttable}
\end{table}

\subsection{Cluster parameters}
\label{sect:pop_syn:results}

For various combinations of $n_\star$ and $\renc$ (considered as fixed grid parameters), we simulate $N_\mathrm{MC}=2000$ systems for a time $\tau_\mathrm{age}=10\,\mathrm{Gyr}$, with different initial planetary eccentricities and random seeds for the encounters. The adopted values of $\renc$ are 25, 50, 75, and 100 $\au$. An overview of the initial conditions is given in Table~\ref{table:par}.

\begin{figure}
\center
\iftoggle{ApJFigs}{
\includegraphics[scale = 0.48, trim = 10mm -5mm 0mm 0mm]{harris.eps}
}{
\includegraphics[scale = 0.48, trim = 10mm -5mm 0mm 0mm]{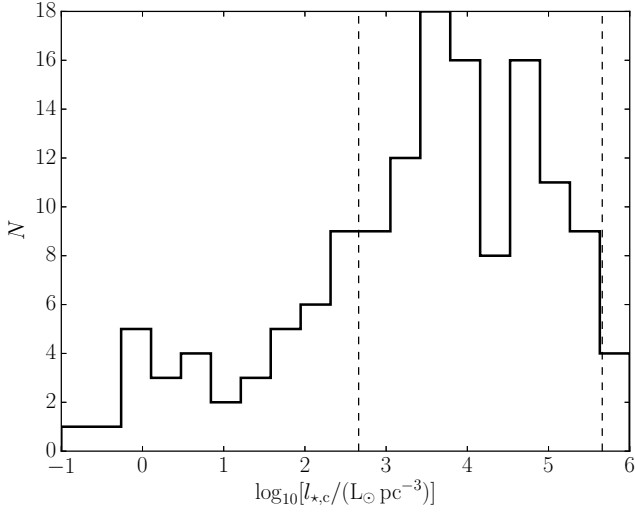}
}
\caption { Distribution of central luminosity densities for the MW GCs in the Harris catalog (\citealt{1996AJ....112.1487H}, 2010 Edition). The assumed range of the luminosity densities in our simulations (Table~\ref{table:par}) is indicated by the two vertical dashed lines. }
\label{fig:harris}
\end{figure}

In particular, we choose the one-dimensional velocity dispersion to be $\sigma=6\,\mathrm{km\,s^{-1}}$, comparable to the typical velocity dispersion of Milky Way (MW) GCs (the mean velocity dispersion in the Harris catalog is $6.3\,\mathrm{km\,s^{-1}}$; \citealt{1996AJ....112.1487H}, 2010 Edition). The number densities $n_\star$ in our simulations range between $10^3$ and $10^{6}$ $\mathrm{pc^{-3}}$. Using the \textsc{SSE} stellar evolution code \citep{2000MNRAS.315..543H} to compute the final masses and luminosities, assuming an age of 5 Gyr and a metallicity of $Z=0.001$, we find that these number densities correspond to mass densities $\rho_\star$ between $2.6\times10^2$ and $2.6 \times 10^5\,\msun\,\mathrm{pc^{-3}}$, and luminosity densities $l_\star$ between $4.5 \times10^2$ and $4.5 \times 10^5 \, \lsun \, \mathrm{pc^{-3}}$. 

In \F\,\ref{fig:harris}, we show the distribution of the central luminosity densities, $l_{\star,\mathrm{c}}$, in the MW GCs (\citealt{1996AJ....112.1487H}; 2010 Edition). The assumed range in our simulations is indicated with the two vertical dashed lines. The MW GC central luminosity densities peak around $10^4 \, \lsun\,\mathrm{pc^{-3}}$, which is near the logarithmic midpoint of our range of simulated densities. We exclude number densities lower than $10^3 \, \mathrm{pc^{-3}}$ even though these densities occur in some MW GCs, because number densities lower than $\sim 10^3 \, \mathrm{pc^{-3}}$ yield few to no migrating planets in our simulations (see \S\,\ref{sect:pop_syn:fractions} below).

\subsection{Encounter properties}
\label{sect:pop_syn:enc_prop}

\begin{figure}
\center
\iftoggle{ApJFigs}{
\includegraphics[scale = 0.48, trim = 10mm 30mm 0mm 40mm]{encounter_properties_run04_enc_b}
}{
\includegraphics[scale = 0.48, trim = 10mm 30mm 0mm 40mm]{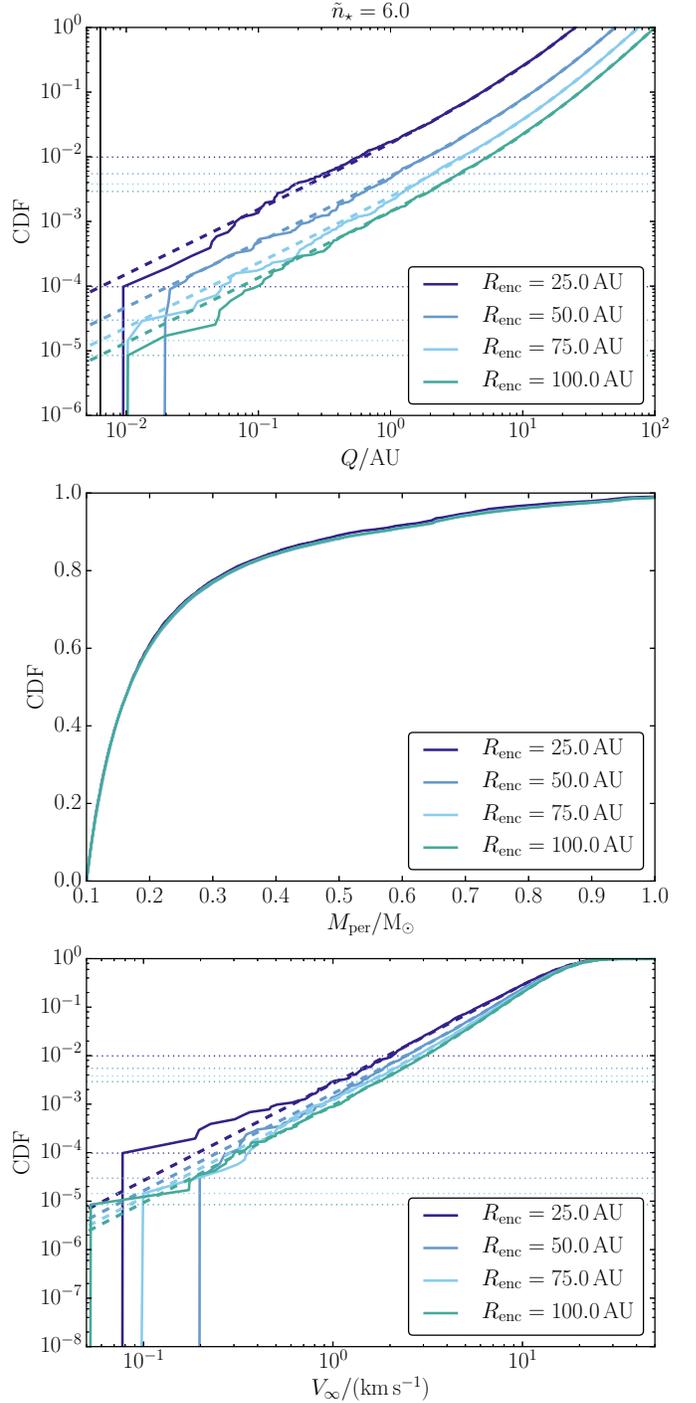}
}
\caption { Encounter properties. Solid colored lines: the cumulative distributions of the periapsis distances $Q$ (top panel), the perturber masses $\mper$ (middle panel), and the speeds at infinity $V_\infty$ (bottom panel), according to the method described in \S\,\ref{sect:pop_syn:enc_gen}. Each color corresponds to a different encounter radius $\renc$, indicated in the legends. In the top panel, the vertical solid black line shows the minimum estimated $Q$ (eq.~\ref{eq:Q_min_est}). In the top and bottom panels, the colored dashed lines show the expected analytic distributions, equations~(\ref{eq:CDF_Q}) and (\ref{eq:CDF_V_inf}), respectively. Also in the top and bottom panels, the upper and lower sets of colored horizontal dotted lines show $1/\sqrt{N_\mathrm{enc}}$ (the Poisson noise in the simulations) and $1/N_\mathrm{enc}$ for each value of $\renc$, where $N_\mathrm{enc}$ is the number of sampled encounters.}
\label{fig:enc}
\end{figure}

Before presenting our main results in \S\,\ref{sect:pop_syn:results}, we briefly discuss properties of the encounters generated via the procedure outlined in \S\,\ref{sect:pop_syn:enc_gen} and compare to analytic results. In \F\,\ref{fig:enc}, we show the cumulative distributions of the periapsis distances $Q$ (top panel), the perturber masses $\mper$ (middle panel), and the speeds at infinity $V_\infty$ (bottom panel), obtained numerically through the method described in \S\,\ref{sect:pop_syn:enc_gen}. In each panel, distributions are shown for the four different encounter sphere radii $\renc$ adopted in the simulations (see Table\,\ref{table:par}) and in all panels $\tilde{n}_\star \equiv \log_{10}(n_\star/\mathrm{pc^{-3}}) = 6$.

The top panel shows the distributions of $Q$. The vertical black solid line shows the expected minimum $Q$ to occur, which can be estimated by setting the collision time for encounters with $r_\mathrm{coll} = Q_\mathrm{min,est}$ equal to the age of the cluster, $\tau_\mathrm{age}$. Following a similar derivation as in \citet[S7.5.8]{2008gady.book.....B} we find
\begin{align}
\label{eq:Q_min_est}
\nonumber Q_\mathrm{min,est} &= -\frac{G( \langle \mper \rangle + M_\star)}{2\srel^2} + \biggl [ \left (\frac{G( \langle \mper \rangle + M_\star)}{2\srel^2} \right )^2 \\
&\quad + \left ( 2 \sqrt{2\pi} \, n_\star \srel \tau_\mathrm{age} \right )^{-1} \biggl ]^{1/2},
\end{align}
where $\langle \mper \rangle$ is the mean perturber mass (in the simulations, $\langle \mper \rangle \simeq 0.26\, \msun$). Note that in the absence of gravitational focusing, this reduces to $Q_\mathrm{min,est} = \left ( 2 \sqrt{2\pi} \, n_\star \srel \tau_\mathrm{age} \right )^{-1/2}$, which is also easily obtained from equation~(\ref{eq:Gamma_limit}) by setting $1 = \Gamma \tau_\mathrm{age}$ with $\renc = Q_\mathrm{min,est}$. The simulated encounters satisfy $\mathrm{min}(Q) > Q_\mathrm{min,est}$. Note that $\mathrm{min}(Q)$ is determined by a single encounter, and therefore it fluctuates among the realizations with different $\renc$. The upper colored horizontal dotted lines show an estimate of the noise level in the CDF, $1/\sqrt{N_\mathrm{enc}}$ where $N_\mathrm{enc}$ is the number of sampled encounters; the lower colored horizontal dotted lines show $1/N_\mathrm{enc}$.

The colored dashed lines in the top panel of \F\,\ref{fig:enc} show the expected cumulative $Q$ distribution, 
\begin{align}
\label{eq:CDF_Q}
\mathrm{CDF}(<Q) \propto Q^2 + \frac{G( \langle \mper \rangle + M_\star) Q }{\srel^2}.
\end{align}
(This can be derived from eq.~\ref{eq:Q_min_est} by solving for $\tau_\mathrm{age}^{-1}$ and noting that the CDF out to some periapsis distance $Q$ must be proportional to $1/\tau_\mathrm{age}$ for $Q_\mathrm{min,est}=Q$.) Equation~(\ref{eq:CDF_Q}) is normalized by the condition that the CDF is unity at $Q=\renc$, since this is the largest possible sampled $Q$. The simulated encounters (solid colored lines) are consistent with equation~(\ref{eq:CDF_Q}), although there is noticeable noise for $Q\lesssim 10^{-1} \, \au$ because of the small number of close encounters. 

The middle panel of \F\,\ref{fig:enc} shows the distribution of the perturber masses. The median perturber mass is $\simeq 0.17\,\msun$, and the mean is $\simeq 0.26 \,\msun$. Despite gravitational focusing, there is little observable dependence of the mass distributions on $\renc$ in these plots, largely because the effects of focusing are strongest for the small fraction of stars with the largest masses. 

Finally, the bottom panel of \F\,\ref{fig:enc} shows the distribution of the perturber speeds at infinity. The colored dashed lines show the expected cumulative $V_\infty$ distribution, 
\begin{align}
\label{eq:CDF_V_inf}
\nonumber &\mathrm{CDF}(<V_\infty) = 1 - \exp \left ( -\frac{V_\infty^2}{2\srel^2} \right ) \\
&\quad \times \left[ 1 + \frac{\renc V_\infty^2}{2\srel^2 \renc + 2 G (M_\star + \langle \mper \rangle )} \right ].
\end{align}
The sampled $V_\infty$'s are consistent with equation~(\ref{eq:CDF_V_inf}), although there is noticeable noise for $V_\infty\lesssim 1\,\mathrm{km\,s^{-1}}$. The median sampled $V_\infty$ depends weakly on $\renc$, varying from $\simeq 13.5\, \mathrm{km\,s^{-1}}$ for $\renc=25\,\au$ to $\simeq 14.9\, \mathrm{km\,s^{-1}}$ for $\renc=100\,\au$.

\subsection{Stopping conditions}
\label{sect:pop_syn:sc}
In the simulations, we distinguish among the following six outcomes (see \S\,\ref{sect:pop_syn:example} below for examples):
\begin{enumerate}

\item {\it HJ formation}: the planetary orbit shrinks to an orbital period that lies between 1 and 10 d, due to encounters, tidal dissipation, or a combination of the two. If the orbit is circularized ($e<10^{-3}$), we immediately stop the integration --- there could be further perturbations due to tidal forces from encounters, but these are extremely weak because the semimajor axis of the orbit is so small. As stated earlier, we neglect stellar tides, so once the orbit is circularized, tidal evolution ceases. In some cases, after $\tau_\mathrm{age}$ the orbital period  lies between 1 and 10 d but the orbit is not yet fully circularized. In this case, we also consider the planet to have become an HJ, unless $a(1-e)<r_\mathrm{t}$ (see below).

\item {\it Warm Jupiter (WJ) formation}: the planetary orbit has shrunk after $\tau_\mathrm{age}$ to an orbital period between 10 and 100 d. The planetary orbit may still be (moderately) eccentric.

\item {\it Tidal disruption}: the planet-host star separation has reached an instantaneous value $r < r_\mathrm{t}$, where the tidal disruption radius $r_\mathrm{t}$ is assumed to be
\begin{align}
\label{eq:r_t}
r_\mathrm{t} = \eta R_\mathrm{p} \left (\frac{ M_\star }{M_\mathrm{p}} \right )^{1/3}.
\end{align}
Here, $\eta$ is a dimensionless parameter, which we assume to be $\eta=2.7$ \citep{2011ApJ...732...74G}. For a solar-mass host star and a planet of Jupiter's mass and radius, which are assumed here, $r_\mathrm{t}\simeq0.013\,\au$. We do not consider collisions with the perturber as a separate outcome, although a fraction of the ``tidal disruption'' outcomes will in fact be collisions with the host star.

\item {\it Unbinding}: after an encounter, the semimajor axis of the planet with respect to its host star is negative or larger than $10\,\au$. In the latter case, the planet is technically still bound to its host star. However, the wide orbit makes the planet extremely susceptible to future encounters even for the lowest densities that we consider, very likely leading to ejection. 

\item {\it Transfer}: the planet is captured by a perturber, which we consider to be the case if the semimajor axis of the planet with respect to the perturber after the encounter is positive and less than $5\,\au$. In principle, a new simulation could be started with the captured planet around the last perturber, subject to further perturbations from encounters that could, e.g., lead to HJ formation. However, this further investigation is beyond the ambitions of this paper. 

\item {\it No migration}: none of the above occurred. Nevertheless, the semimajor axis and/or eccentricity may still have been affected significantly.

\end{enumerate}

\subsection{Simulation results}
\label{sect:pop_syn:results}

\subsubsection{Examples}
\label{sect:pop_syn:example}

\begin{figure*}
\center
\iftoggle{ApJFigs}{
\includegraphics[scale = 0.4, trim = 35mm 10mm 0mm 0mm]{examples3_run05_ex.eps}
}{
\includegraphics[scale = 0.4, trim = 35mm 10mm 0mm 0mm]{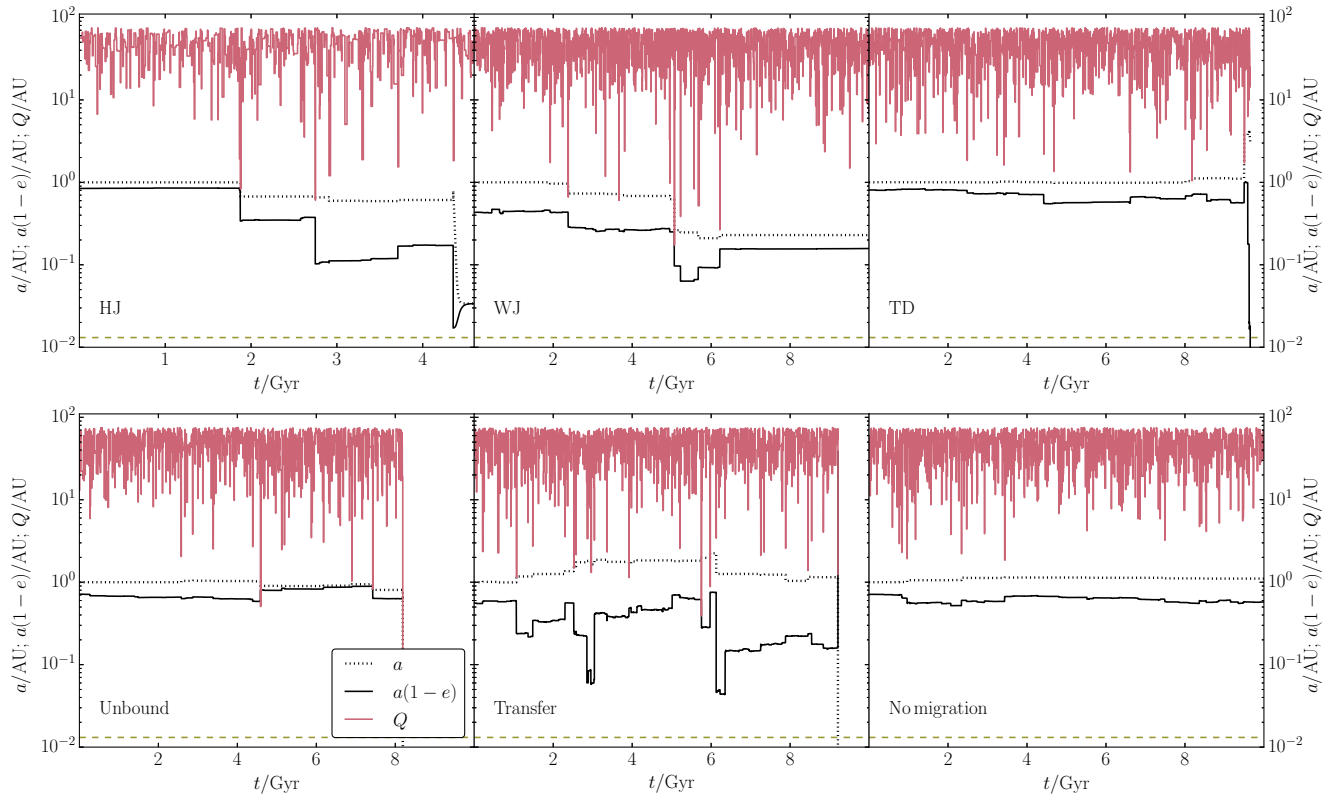}
}
\caption { Six examples to illustrate the typical eccentricity and semimajor axis evolution for the outcomes defined in \S\,\ref{sect:pop_syn:sc}. In these examples, the initial planetary semimajor axis is $1\,\au$, the encounter radius is $\renc = 75\,\au$, and the stellar density is $n_\star = 2 \times 10^4 \, \mathrm{pc^{-3}}$. The vertical axes show the semimajor axis $a$ (black dotted lines), the periapsis distance $a(1-e)$ (black solid lines), and the encounter periapsis distances $Q$ (red solid lines). Top row from left to right: HJ formation, WJ formation, tidal disruption; bottom row: unbound planet, transferred planet, and no migration. In each panel, the yellow horizontal dashed line shows the tidal disruption radius (eq.~\ref{eq:r_t}). Note that each change in the height of the red solid lines corresponds to a new encounter with a certain $Q$; the horizontal sections indicate the time between encounters, not the passage time of the encounters themselves. The ratio of the passage time to the time between encounters for the chosen parameters is $\approx 4.8 \times 10^{-6}$ (see equation~\ref{eq:ratio_t_pas_t_enc}). }
\label{fig:examples}
\end{figure*}

In Fig.~\ref{fig:examples}, we show a number of examples to illustrate the typical eccentricity and semimajor axis evolution for the outcomes defined in \S\,\ref{sect:pop_syn:sc}. In these examples, the initial planetary semimajor axis is $1\,\au$, the encounter radius is $\renc = 75\,\au$, and the stellar density is $n_\star = 2 \times 10^4 \, \mathrm{pc^{-3}}$. 

In the top-left panel, an HJ is formed within $\approx 4.5\, \mathrm{Gyr}$ of evolution. After $1.8\,\mathrm{Gyr}$, a number of strong perturbers with a $Q$ of a few $\au$ gradually decrease the periapsis distance, while leaving the semimajor axis relatively unchanged. After a strong encounter at $\approx 4.4 \, \mathrm{Gyr}$, the periapsis distance is reduced to a small enough value that strong tidal evolution is triggered. The orbit circularizes at a semimajor axis of $\approx 0.03\,\au$. 
In the top-middle panel, a WJ is formed through a series of encounters in which the semimajor axis decreases to $\approx 0.2\,\au$ and the periapsis distance decreases to $\approx 0.16\,\au$. The planet survives for 10 Gyr, and has final orbital properties characteristic of WJs ($a\simeq 0.23\,\au$; $e \simeq 0.31$). Tidal dissipation played no role in this process.
The top-right panel shows an example of tidal disruption; the semimajor axis is not much affected by encounters, whereas the periapsis distance is gradually decreased. A single encounter at $\approx 9.5\,\mathrm{Gyr}$ increases the semimajor axis to $\approx 4.3\,\au$, making the planet more susceptible to further perturbations, and it is disrupted shortly thereafter. 

In the bottom-left panel, a strong encounter with $Q\approx 0.12 \, \au$ unbinds the planet at $\approx 8.2\,\mathrm{Gyr}$. Another strong encounter occurs at $\approx 9.2\,\mathrm{Gyr}$ in the bottom-middle panel, and the planet is transferred to the perturber. Lastly, in the bottom-right panel, no destructive or migration-inducing encounters occur, and the planet survives without significant tidal dissipation, although the eccentricity has been slightly excited and the semimajor axis slightly increased.

\subsubsection{Simulation outcome fractions}
\label{sect:pop_syn:fractions}

\begin{figure*}
\center
\iftoggle{ApJFigs}{
\includegraphics[scale = 0.48, trim = 10mm 10mm 10mm 0mm]{fractions_run05.eps}
\includegraphics[scale = 0.48, trim = 10mm 10mm 10mm 0mm]{fractions_run05_a4.eps}
}{
\includegraphics[scale = 0.48, trim = 10mm 10mm 10mm 0mm]{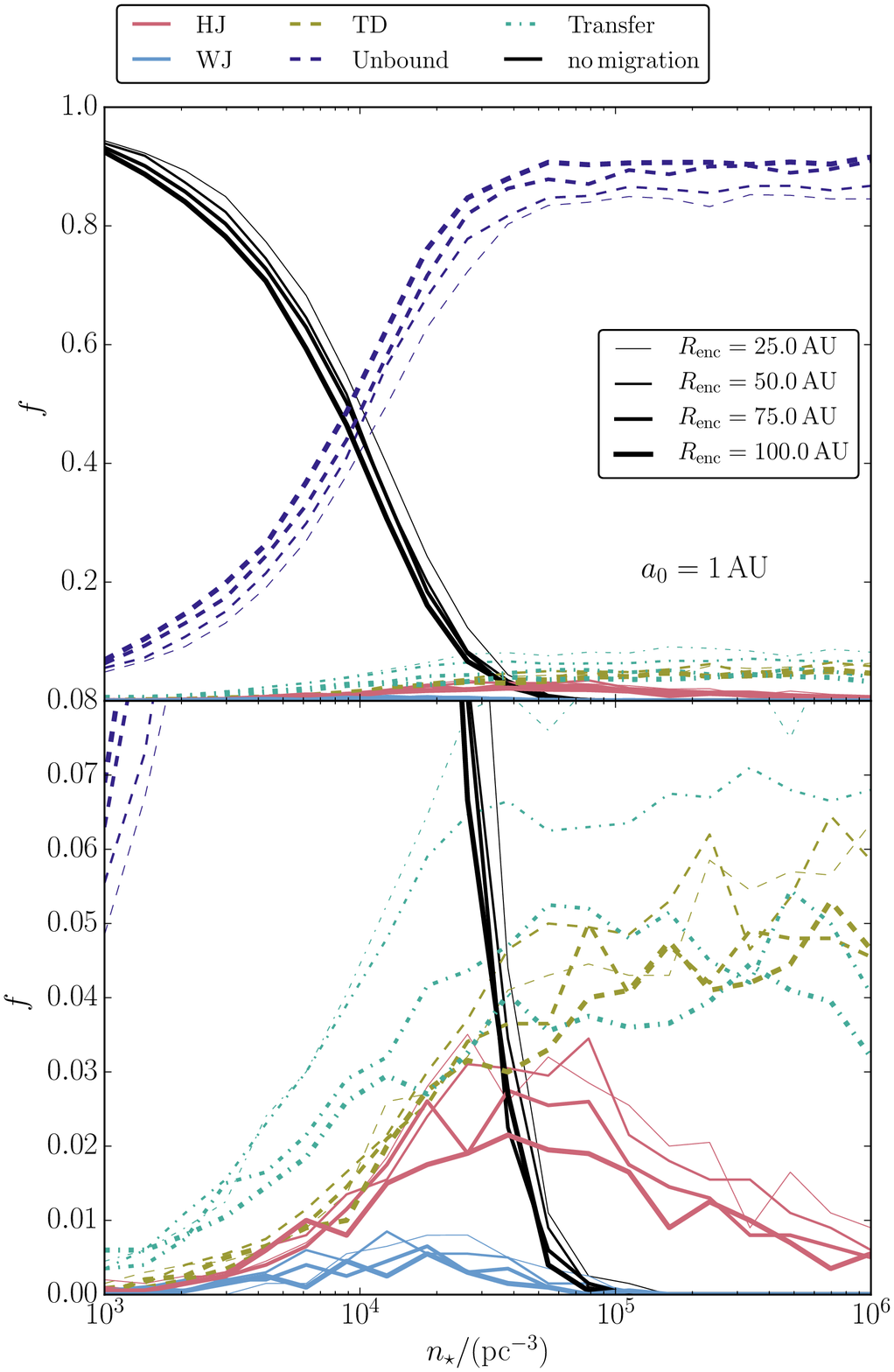}
\includegraphics[scale = 0.48, trim = 10mm 10mm 10mm 0mm]{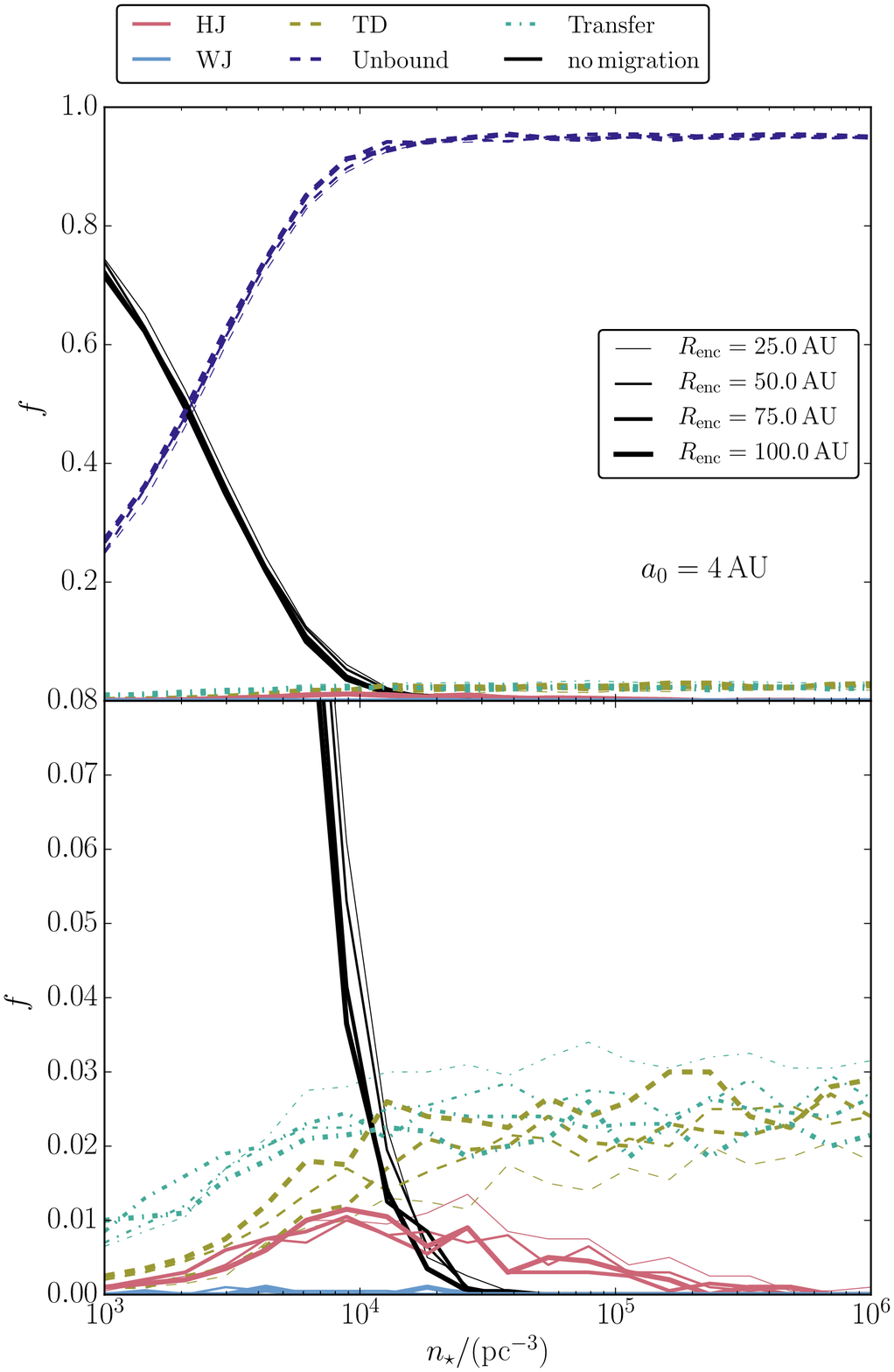}
}
\caption { Fractions of the six outcomes (\S\,\ref{sect:pop_syn:sc}) as a function of the stellar number density $n_\star$. The initial planetary semimajor axis is $a_0 = 1\,\au$ ($a_0=4\,\au$) in the left (right) set of panels. Red solid lines correspond to HJ formation, blue solid lines correspond to WJ formation, yellow dashed lines correspond to tidal disruption events, dark blue dashed lines correspond to unbound planets, green dotted-dashed lines correspond to transferred planets, and solid black lines (`no migration') correspond to planets that remain bound to their host star with periods above the assumed WJ threshold, 100 d (see \S\,\ref{sect:pop_syn:sc}). Results are shown for four values of the encounter sphere radius $\renc$; line thickness increases with increasing $\renc$. The bottom panels are zoomed-in versions of the top panels, showing in more detail the less common outcomes. The data (for $a_0=1\,\au$) are also given in Table\,\ref{table:outcomes}. }
\label{fig:fractions}
\end{figure*}

\begin{table}
{\fontsize{7.5}{7} \selectfont
\begin{tabular}{cccccccc}
\toprule
$\tilde{n}_\star$ & $\renc/\au$ & $f_\mathrm{HJ}$ & $f_\mathrm{WJ}$ & $f_\mathrm{TD}$ & $f_\mathrm{un.}$ & $f_\mathrm{tr.}$ & $f_\mathrm{no\,migr.}$ \\
\midrule
3.0 & 25 & 0.00 & 0.00 & 0.00 & 0.05 & 0.00 & 0.94 \\
3.0 & 50 & 0.00 & 0.00 & 0.00 & 0.06 & 0.00 & 0.94 \\
3.0 & 75 & 0.00 & 0.00 & 0.00 & 0.06 & 0.00 & 0.93 \\
3.0 & 100 & 0.00 & 0.00 & 0.00 & 0.07 & 0.01 & 0.92 \\
3.16 & 25 & 0.00 & 0.00 & 0.00 & 0.07 & 0.01 & 0.92 \\
3.16 & 50 & 0.00 & 0.00 & 0.00 & 0.07 & 0.01 & 0.92 \\
3.16 & 75 & 0.00 & 0.00 & 0.00 & 0.09 & 0.00 & 0.90 \\
3.16 & 100 & 0.00 & 0.00 & 0.00 & 0.10 & 0.01 & 0.89 \\
3.32 & 25 & 0.00 & 0.00 & 0.00 & 0.09 & 0.01 & 0.89 \\
3.32 & 50 & 0.00 & 0.00 & 0.00 & 0.11 & 0.01 & 0.87 \\
3.32 & 75 & 0.00 & 0.00 & 0.00 & 0.13 & 0.01 & 0.86 \\
3.32 & 100 & 0.00 & 0.00 & 0.00 & 0.15 & 0.01 & 0.84 \\
3.47 & 25 & 0.00 & 0.00 & 0.00 & 0.13 & 0.01 & 0.85 \\
3.47 & 50 & 0.00 & 0.00 & 0.01 & 0.15 & 0.02 & 0.82 \\
3.47 & 75 & 0.00 & 0.00 & 0.00 & 0.17 & 0.01 & 0.80 \\
3.47 & 100 & 0.00 & 0.00 & 0.01 & 0.20 & 0.01 & 0.78 \\
3.63 & 25 & 0.00 & 0.00 & 0.01 & 0.19 & 0.02 & 0.77 \\
3.63 & 50 & 0.01 & 0.00 & 0.01 & 0.21 & 0.02 & 0.74 \\
3.63 & 75 & 0.00 & 0.00 & 0.01 & 0.24 & 0.02 & 0.73 \\
3.63 & 100 & 0.01 & 0.00 & 0.01 & 0.26 & 0.01 & 0.71 \\
3.79 & 25 & 0.01 & 0.00 & 0.01 & 0.27 & 0.03 & 0.68 \\
3.79 & 50 & 0.01 & 0.01 & 0.01 & 0.30 & 0.03 & 0.65 \\
3.79 & 75 & 0.01 & 0.00 & 0.01 & 0.33 & 0.02 & 0.63 \\
3.79 & 100 & 0.01 & 0.00 & 0.01 & 0.37 & 0.02 & 0.59 \\
3.95 & 25 & 0.01 & 0.01 & 0.01 & 0.38 & 0.04 & 0.55 \\
3.95 & 50 & 0.01 & 0.00 & 0.02 & 0.41 & 0.04 & 0.52 \\
3.95 & 75 & 0.01 & 0.00 & 0.01 & 0.44 & 0.03 & 0.50 \\
3.95 & 100 & 0.01 & 0.00 & 0.01 & 0.49 & 0.03 & 0.47 \\
4.11 & 25 & 0.02 & 0.01 & 0.03 & 0.50 & 0.05 & 0.40 \\
4.11 & 50 & 0.02 & 0.01 & 0.02 & 0.57 & 0.05 & 0.34 \\
4.11 & 75 & 0.02 & 0.00 & 0.02 & 0.58 & 0.03 & 0.34 \\
4.11 & 100 & 0.01 & 0.00 & 0.02 & 0.63 & 0.03 & 0.31 \\
4.26 & 25 & 0.03 & 0.01 & 0.03 & 0.63 & 0.06 & 0.24 \\
4.26 & 50 & 0.02 & 0.01 & 0.03 & 0.68 & 0.06 & 0.20 \\
4.26 & 75 & 0.03 & 0.01 & 0.03 & 0.72 & 0.04 & 0.18 \\
4.26 & 100 & 0.02 & 0.01 & 0.03 & 0.76 & 0.03 & 0.16 \\
4.42 & 25 & 0.04 & 0.01 & 0.03 & 0.72 & 0.08 & 0.12 \\
4.42 & 50 & 0.03 & 0.01 & 0.04 & 0.78 & 0.06 & 0.08 \\
4.42 & 75 & 0.02 & 0.00 & 0.03 & 0.82 & 0.04 & 0.08 \\
4.42 & 100 & 0.02 & 0.00 & 0.03 & 0.85 & 0.03 & 0.07 \\
4.58 & 25 & 0.03 & 0.01 & 0.04 & 0.80 & 0.08 & 0.04 \\
4.58 & 50 & 0.03 & 0.01 & 0.05 & 0.82 & 0.07 & 0.03 \\
4.58 & 75 & 0.03 & 0.00 & 0.04 & 0.86 & 0.05 & 0.02 \\
4.58 & 100 & 0.02 & 0.00 & 0.03 & 0.88 & 0.04 & 0.03 \\
4.74 & 25 & 0.03 & 0.00 & 0.04 & 0.83 & 0.08 & 0.01 \\
4.74 & 50 & 0.03 & 0.00 & 0.05 & 0.85 & 0.06 & 0.01 \\
4.74 & 75 & 0.03 & 0.00 & 0.04 & 0.88 & 0.05 & 0.01 \\
4.74 & 100 & 0.02 & 0.00 & 0.03 & 0.91 & 0.04 & 0.00 \\
4.89 & 25 & 0.03 & 0.00 & 0.04 & 0.84 & 0.08 & 0.00 \\
4.89 & 50 & 0.03 & 0.00 & 0.05 & 0.85 & 0.06 & 0.00 \\
4.89 & 75 & 0.03 & 0.00 & 0.05 & 0.87 & 0.05 & 0.00 \\
4.89 & 100 & 0.02 & 0.00 & 0.04 & 0.90 & 0.04 & 0.00 \\
5.05 & 25 & 0.03 & 0.00 & 0.04 & 0.85 & 0.08 & 0.00 \\
5.05 & 50 & 0.02 & 0.00 & 0.05 & 0.87 & 0.06 & 0.00 \\
5.05 & 75 & 0.02 & 0.00 & 0.04 & 0.89 & 0.05 & 0.00 \\
5.05 & 100 & 0.02 & 0.00 & 0.04 & 0.91 & 0.04 & 0.00 \\
5.21 & 25 & 0.02 & 0.00 & 0.04 & 0.85 & 0.09 & 0.00 \\
5.21 & 50 & 0.02 & 0.00 & 0.05 & 0.86 & 0.07 & 0.00 \\
5.21 & 75 & 0.01 & 0.00 & 0.05 & 0.89 & 0.05 & 0.00 \\
5.21 & 100 & 0.01 & 0.00 & 0.05 & 0.91 & 0.04 & 0.00 \\
5.37 & 25 & 0.02 & 0.00 & 0.06 & 0.83 & 0.09 & 0.00 \\
5.37 & 50 & 0.02 & 0.00 & 0.06 & 0.86 & 0.07 & 0.00 \\
5.37 & 75 & 0.01 & 0.00 & 0.04 & 0.90 & 0.04 & 0.00 \\
5.37 & 100 & 0.01 & 0.00 & 0.04 & 0.91 & 0.04 & 0.00 \\
5.53 & 25 & 0.01 & 0.00 & 0.05 & 0.85 & 0.08 & 0.00 \\
5.53 & 50 & 0.02 & 0.00 & 0.05 & 0.87 & 0.07 & 0.00 \\
5.53 & 75 & 0.01 & 0.00 & 0.05 & 0.90 & 0.04 & 0.00 \\
5.53 & 100 & 0.01 & 0.00 & 0.04 & 0.90 & 0.04 & 0.00 \\
5.68 & 25 & 0.02 & 0.00 & 0.06 & 0.85 & 0.07 & 0.00 \\
5.68 & 50 & 0.01 & 0.00 & 0.05 & 0.87 & 0.07 & 0.00 \\
5.68 & 75 & 0.01 & 0.00 & 0.05 & 0.89 & 0.05 & 0.00 \\
5.68 & 100 & 0.01 & 0.00 & 0.04 & 0.91 & 0.04 & 0.00 \\
5.84 & 25 & 0.01 & 0.00 & 0.06 & 0.85 & 0.09 & 0.00 \\
5.84 & 50 & 0.01 & 0.00 & 0.06 & 0.86 & 0.07 & 0.00 \\
5.84 & 75 & 0.01 & 0.00 & 0.05 & 0.90 & 0.05 & 0.00 \\
5.84 & 100 & 0.00 & 0.00 & 0.05 & 0.90 & 0.04 & 0.00 \\
6.0 & 25 & 0.01 & 0.00 & 0.06 & 0.85 & 0.08 & 0.00 \\
6.0 & 50 & 0.01 & 0.00 & 0.06 & 0.87 & 0.07 & 0.00 \\
6.0 & 75 & 0.01 & 0.00 & 0.05 & 0.91 & 0.04 & 0.00 \\
6.0 & 100 & 0.01 & 0.00 & 0.05 & 0.92 & 0.03 & 0.00 \\
\bottomrule
\end{tabular}
}
\caption{ Simulated outcome fractions for the six outcomes listed in \S\,\ref{sect:pop_syn:sc} ($f_\mathrm{un.}$: unbound; $f_\mathrm{tr.}$: transferred; $f_\mathrm{no\,migr.}$: no migration). The same data are also shown in \F\,\ref{fig:fractions}. We define $\tilde{n}_\star \equiv \log_{10}(n_\star/\mathrm{pc^{-3}})$. The statistical uncertainty in the outcome fractions is about 0.02.}
\label{table:outcomes}
\end{table}

\F\,\ref{fig:fractions} shows the fractions of the six outcomes (see \S\,\ref{sect:pop_syn:sc}) as a function of the stellar number density $n_\star$. The initial planetary semimajor axis is $a_0 = 1\,\au$ ($a_0=4\,\au$) in the left (right) set of panels. In \F\,\ref{fig:fractions} and in subsequent figures, red solid lines correspond to HJ formation, blue solid lines correspond to WJ formation, yellow dashed lines correspond to tidal disruption events, dark blue dashed lines correspond to unbound planets, green dotted-dashed lines correspond to transferred planets, and solid black lines correspond to nonmigrating planets. The bottom panels are zoomed-in versions of the top panels, showing in more detail the less common outcomes. The expected Poisson fluctuation in the outcome fractions $f$ from our $N_\mathrm{MC}=2000$ realizations is $1/\sqrt{N_\mathrm{MC}}\simeq 0.02$. The data (for $a_0=1\,\au$) are also given in Table\,\ref{table:outcomes}. 

We first discuss the case $a_0=1\,\au$. For the lowest densities, the encounter rate is too low to strongly affect the planetary orbit, and the fraction of nonmigrating planets is near 100\%. As $n_\star$ is increased, the nonmigrating fraction decreases smoothly to zero near $n_\star= 10^5\, \mathrm{pc^{-3}}$, while the fraction of unbound planets increases to $\approx 0.9$ at the same density and stays near that value for even higher densities. The migration fraction, i.e., the sum of $f_\mathrm{HJ}$, $f_\mathrm{WJ}$ and $f_\mathrm{TD}$, is nearly zero for $n_\star = 10^3\,\mathrm{pc^{-3}}$, increases to $\approx 0.06$ near $n_\star = 10^5\,\mathrm{pc^{-3}}$, and remains approximately constant at higher densities. Note that the sum of the migration fraction and the nonmigration fraction is not unity, because both categories exclude the unbinding and transfer outcomes. 

The fraction of HJs peaks at $\approx 0.02$ at $n_\star\approx 4\times 10^4\,\mathrm{pc^{-3}}$ (the fraction approaches $\approx 0.03$ for the smallest encounter radii but these values are less reliable). For higher densities, the HJ fraction decreases, whereas the fraction of tidal disruptions and transferred planets increases. These trends can be understood qualitatively by noting that the density needs to be sufficiently high for strong encounters to decrease the periapsis distance of the planet and trigger tidal migration. However, as the density increases, strong encounters become increasingly likely, and therefore more catastrophic outcomes, like tidal disruption or a transfer to the perturber, start to dominate. 

The fraction of unbound planets is weakly dependent on $\renc$.  Nevertheless, the fractions shown in \F\,\ref{fig:fractions} indicate convergence with respect to $\renc$; in particular, there are only small differences between $\renc=75$ and 100 $\au$. 

Comparing the two cases $a_0=1$ and 4 $\au$, we note that the curves showing the fraction of nonmigrating and unbound planets are shifted to the left (lower $n_\star$) for $a_0=4\,\au$, which can be understood qualitatively because planets with larger semimajor axes are more weakly bound. Similarly to the case $a_0=1\,\au$, the migration fraction is approximately constant at high densities ($n_\star \gtrsim 10^4\,\mathrm{pc^{-3}}$), although the absolute value of the migration fraction, $\approx 0.025$, is lower by a factor of $\sim 2$. Furthermore, the HJ fraction peaks at a lower density, in this case around $10^4\,\mathrm{pc^{-3}}$, with a value of $\approx 0.01$ which is a factor of $\sim 2$ lower than the case $a_0=1\,\au$. 

The fractions for the case $a_0=2\,\au$ (not presented here) show intermediate trends between those depicted in \F\,\ref{fig:fractions}. We conclude that a larger $a_0$ tends to lower the HJ and WJ fractions and, more generally, the fractions of `interesting' outcomes such as tidal disruption and planet transfer. In the remainder of \S\,\ref{sect:pop_syn:results}, we focus mainly on the case $a_0=1\,\au$. Most of the results shown in these sections do not depend strongly on $a_0$.

\subsubsection{Orbital changes}
\label{sect:pop_syn:orb}

\begin{figure}
\center
\iftoggle{ApJFigs}{
\includegraphics[scale = 0.42, trim = 20mm 0mm 0mm 0mm]{a_e_plane_run05.eps}
}{
\includegraphics[scale = 0.42, trim = 20mm 0mm 0mm 0mm]{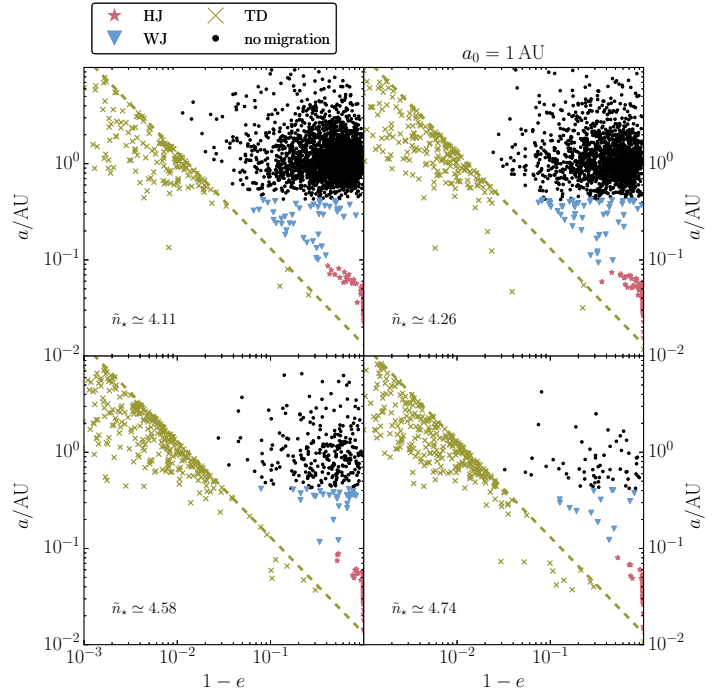}
}
\caption { Orbital elements in the $(a,1-e)$ plane at the end of the simulation, or after a stopping condition occurred, for the HJ (red stars), WJ  (blue triangles), tidal disruption (yellow crosses), and no migration cases (black dots). Each panel corresponds to a different density, indicated therein ($\tilde{n}_\star \equiv \log_{10}[n_\star/\mathrm{pc^{-3}}]$), and the initial semimajor axis is $1\,\au$. The yellow dashed line shows the assumed threshold for tidal disruption, $a(1-e) = r_\mathrm{t}$, with $r_\mathrm{t}$ given by equation~(\ref{eq:r_t}). Transfer outcomes are not shown; see \F\,\ref{fig:transferred_planets} for a similar figure showing the orbital elements of the transferred planets. }
\label{fig:a_e_plane_sim}
\end{figure}

To illustrate how encounters affect the planetary orbit, we show in \F\,\ref{fig:a_e_plane_sim} the orbital elements in the $(a,1-e)$ plane at the end of the simulation, or after a stopping condition occurred, for the outcomes HJ (red stars), WJ  (blue triangles), tidal disruption (yellow crosses), and no migration  (black dots). Each panel corresponds to a different density. In all panels $a_0=1\,\au$. We recall that the initial eccentricities were sampled from a Rayleigh distribution with an rms of 0.33 and maximum value of 0.6 (see \S\,\ref{sect:pop_syn:dyn}). 

The yellow dashed line shows the assumed threshold for tidal disruption, $a(1-e) = r_\mathrm{t}$, with $r_\mathrm{t}$ given by equation~(\ref{eq:r_t}). Obviously, the points corresponding to the tidal disruption outcome lie below this line. Most tidal disruptions occur with relatively large semimajor axes, $a \gtrsim 0.5\,\au$, implying that the encounters leading to tidal disruptions mainly did so by exciting high eccentricities as opposed to strongly decreasing the semimajor axes.

The HJs populate the lower-right part of the $(a,1-e)$ plane. They tend to be either completely circular or have a small eccentricity. Their semimajor/orbital period distribution is discussed in more detail below. The WJs lie in between the HJs and nonmigrating planets, and are substantially more eccentric than the HJs (WJs around field stars in the solar neighborhood also tend to be eccentric, but probably for different reasons). The nonmigrating planets populate a large region in the $(a,1-e)$ diagram, even though they all began with a common semimajor axis of $1\,\au$ and small eccentricities.  Encounters change the semimajor axes, typically by a factor of a few (note that we consider planets to be unbound if $a>10\au$), and can excite the eccentricities to values as high as $\sim 0.99$ (or even higher, at which point the planets become HJs or are tidally disrupted). 

\begin{figure}
\center
\iftoggle{ApJFigs}{
\includegraphics[scale = 0.45, trim = 10mm 5mm 0mm 0mm]{semimajor_axes_run05.eps}
\includegraphics[scale = 0.45, trim = 10mm 5mm 0mm 0mm]{semimajor_axes_run05_a4.eps}
}{
\includegraphics[scale = 0.45, trim = 10mm 5mm 0mm 0mm]{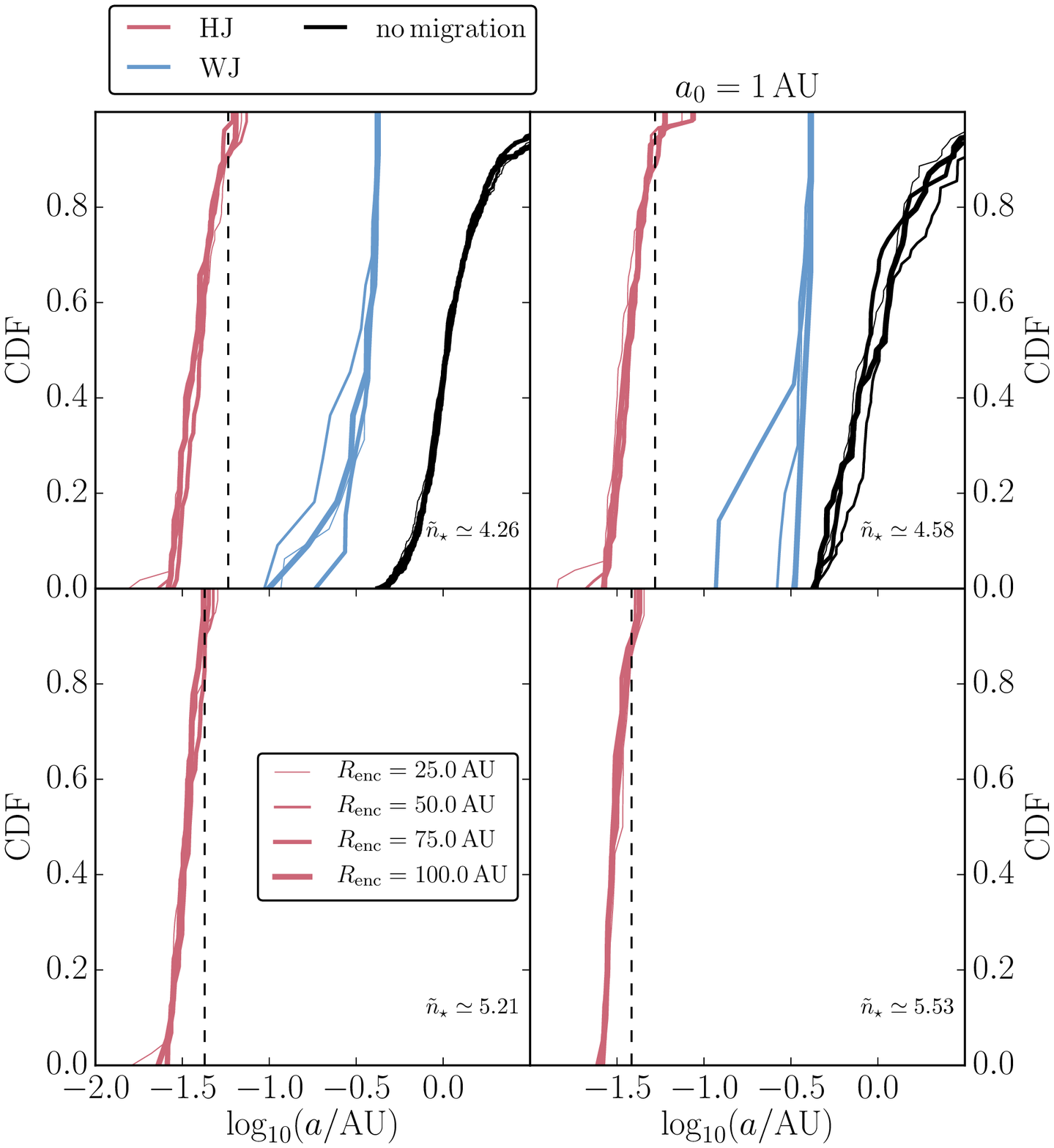}
\includegraphics[scale = 0.45, trim = 10mm 5mm 0mm 0mm]{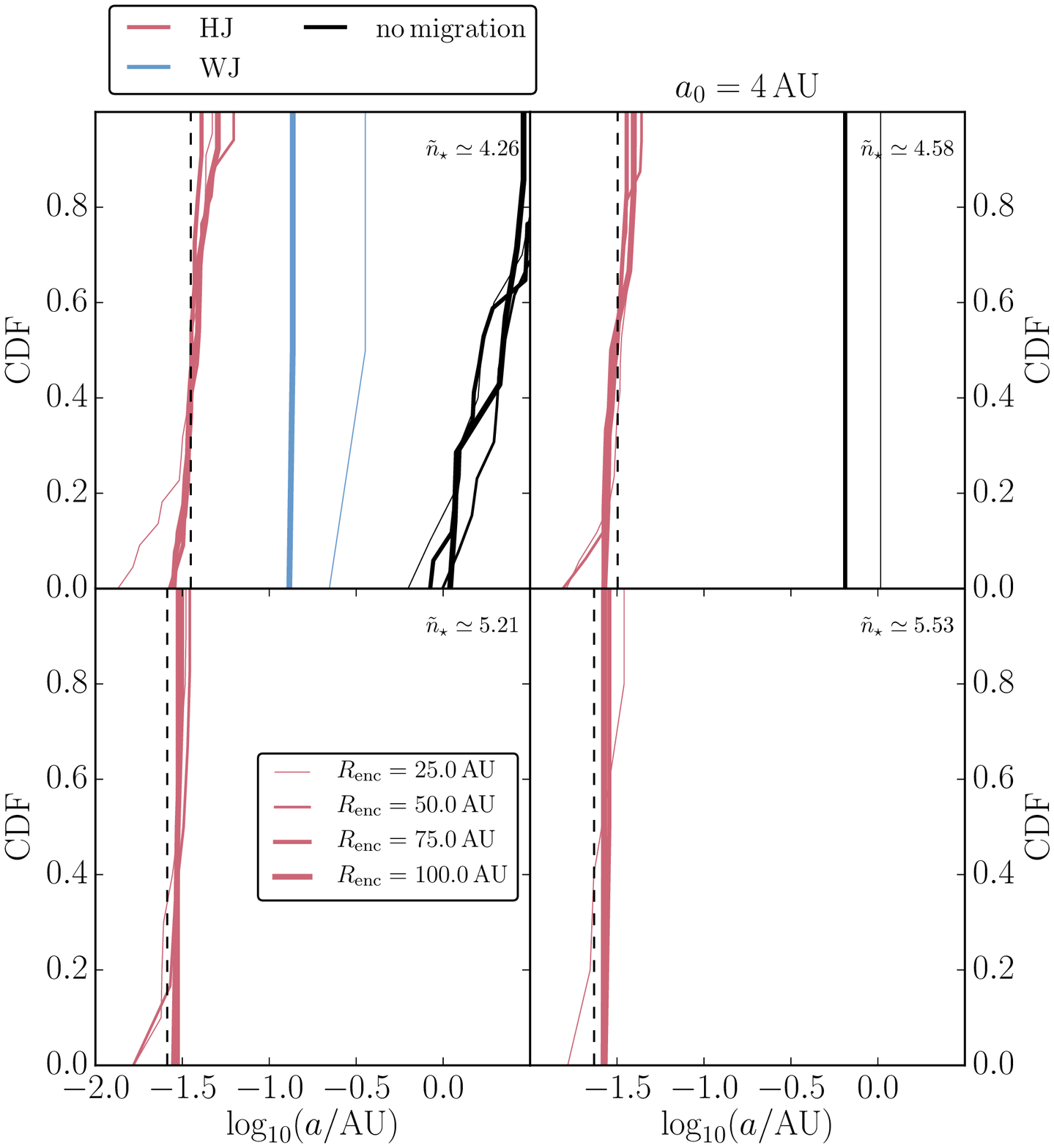}
}
\caption { Cumulative distributions of the semimajor axes separately for the three `nondisruptive' outcomes, i.e., HJ and WJ formation and no migration, for four values of $n_\star$ and various values of $\renc$. The top (bottom) set of panels corresponds to $a_0=1\,\au$ ($a_0=4\,\au$). Increasing line widths correspond to larger $\renc$. Note that not all outcomes occur at all densities. In all panels, the black vertical dashed lines show predictions from the analytic model discussed in \S\,\ref{sect:pop_syn:an}.}
\label{fig:HJ_smas}
\end{figure}

\begin{figure}
\center
\iftoggle{ApJFigs}{
\includegraphics[scale = 0.45, trim = 10mm 0mm 0mm 0mm]{HJ_periods_obs_run05.eps}
}{
\includegraphics[scale = 0.45, trim = 10mm 0mm 0mm 0mm]{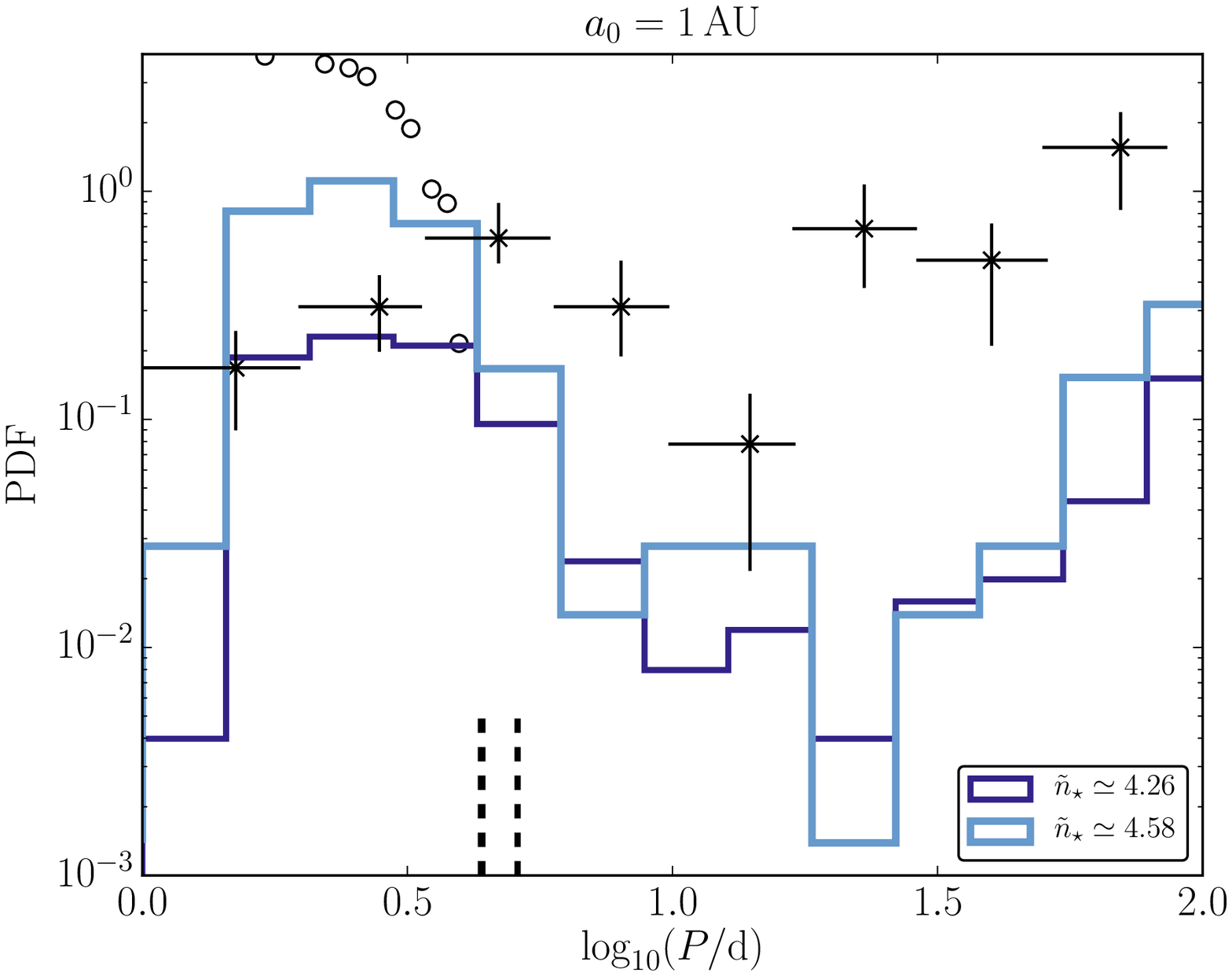}
}
\caption { Colored lines: the orbital period distributions for the nondisruptive outcomes in the simulations (HJ and WJ formation and no migration, combining data from all values of $\renc$, and $a_0=1\,\au$). Black crosses and error bars: the observed distribution for giant planets in the solar neighborhood \citep{2016A&A...587A..64S}. Black open circles: the distribution from simulations by \citet{2016MNRAS.456.3671A} of high-$e$ HJ formation in stellar binaries in the field (data are shown from Fig.~23 of \citealt{2016MNRAS.456.3671A} for $M_\mathrm{p} = 1 M_\mathrm{J}$ and $\chi = 100$, where $\chi \equiv 10 \, \tau_\mathrm{p}/\mathrm{s}$ and $\tau_\mathrm{p}$ is the planetary tidal time lag; with $k_\mathrm{AM,\mathrm{p}} = 0.25$, $M_\mathrm{p} = 1\,M_\mathrm{J}$, and $R_\mathrm{p} = 1 \, R_\mathrm{J}$, $\chi=100$ or $\tau_\mathrm{p} = 10 \, \mathrm{s}$ corresponds to a viscous timescale of $\simeq 0.082 \, \mathrm{yr}$). The distributions are normalized to unit total area. The short black vertical dashed lines show the predictions from the analytic model (\S\,\ref{sect:pop_syn:an}).}
\label{fig:HJ_periods_obs}
\end{figure}

In \F\,\ref{fig:HJ_smas}, we show the cumulative distributions of the semimajor axes separately for the ``nondisruptive'' cases, i.e., HJ and WJ formation and no migration, for four values of $n_\star$ and the various values of $\renc$. The top (bottom) set of panels corresponds to $a_0=1\,\au$ ($a_0=4\,\au$). We repeat that we did not include stellar tides in the simulations, which could affect the HJ semimajor axis distribution if the stellar tides are efficient (in particular, if the star has a convective layer). There is little dependence of the semimajor axis distributions on $\renc$, which provides reassurance that our choices of $\renc$ are large enough. The HJ semimajor axes tend to be slightly smaller for higher densities and larger $a_0$. These trends can be explained with an analytic model that is described below in \S\,\ref{sect:pop_syn:an}. The predictions from that model are shown in \F\,\ref{fig:HJ_smas} with the vertical black dashed lines. 

In \F\,\ref{fig:HJ_periods_obs}, we show the distributions of the final orbital periods for $a_0=1\,\au$. The data have been combined for all three nondisruptive cases and all four values of $\renc$. The different colored lines correspond to two values of $n_\star$. For comparison, we also show with black crosses and error bars the observed distribution for giant planets in the solar neighborhood by \citet{2016A&A...587A..64S}, and, with black open circles, the distribution from simulations by \citet{2016MNRAS.456.3671A} of HJ formation by high-eccentricity migration in stellar binaries. 

Our simulations predict a peak in the HJ period distribution in dense clusters at $\sim$ 2.5 day, which is similar to the predicted peak for stellar binaries in the field, but shorter than the observed peak for field HJs at $\sim$ 5 day. The shorter orbital periods in our simulations and those of \citet{2016MNRAS.456.3671A} compared to the observations are typical for high-$e$ migration scenarios. Note that the final orbital period depends sensitively on the planetary radius (the final semimajor axis is approximately proportional to $R_\mathrm{p}$, see \S\,\ref{sect:pop_syn:an} below), and we assumed a single planetary radius ($R_\mathrm{p} = 1 \, R_\mathrm{J}$). Therefore, our simulations can be made to be more consistent with the observations by, e.g., assuming inflated HJs (e.g., \citealt{2011ApJ...729L...7L}).

The occurrence of WJs in our simulations is a factor of a few lower than HJs (see \F\,\ref{fig:fractions}). In contrast, the observations of \citet{2016A&A...587A..64S} for the solar neighborhood indicate that a large fraction of planets have orbital periods characteristic of WJs. Simulations of high-$e$ migration in stellar binaries (e.g,. \citealt{2016ApJ...829..132P,2016AJ....152..174A}), multiplanet systems \citep{2017MNRAS.464..688H}, and stellar triples \citep{2017MNRAS.466.4107H} also produce few WJs, or a number of WJs inconsistent with observations. Nevertheless, the WJ fractions in our simulations are higher than, e.g., in \citet{2017MNRAS.464..688H} and \citep{2017MNRAS.466.4107H}. This is because the WJs in our simulations tend to be produced directly by encounters without substantial tidal evolution, whereas in purely secular high-$e$ migration, the semimajor axis can only change due to tidal interactions.

\subsubsection{Obliquities}
\label{sect:pop_syn:obl}

\begin{figure}
\center
\iftoggle{ApJFigs}{
\includegraphics[scale = 0.45, trim = 10mm 5mm 0mm 0mm]{stellar_obliquities_run05.eps}
\includegraphics[scale = 0.45, trim = 10mm 0mm 0mm 0mm]{stellar_obliquities_CDF_run05.eps}
}{
\includegraphics[scale = 0.45, trim = 10mm 5mm 0mm 0mm]{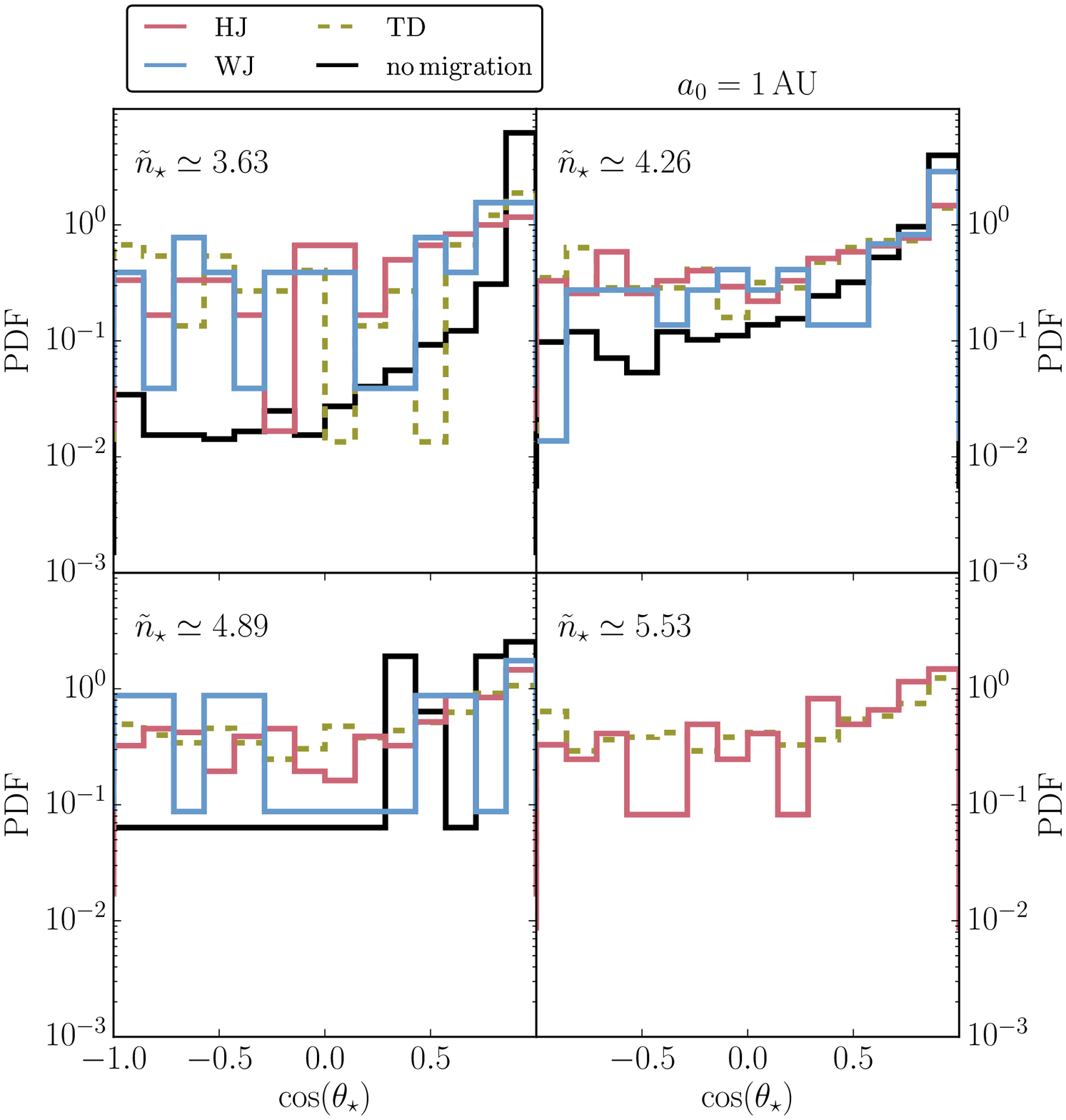}
\includegraphics[scale = 0.45, trim = 10mm 0mm 0mm 0mm]{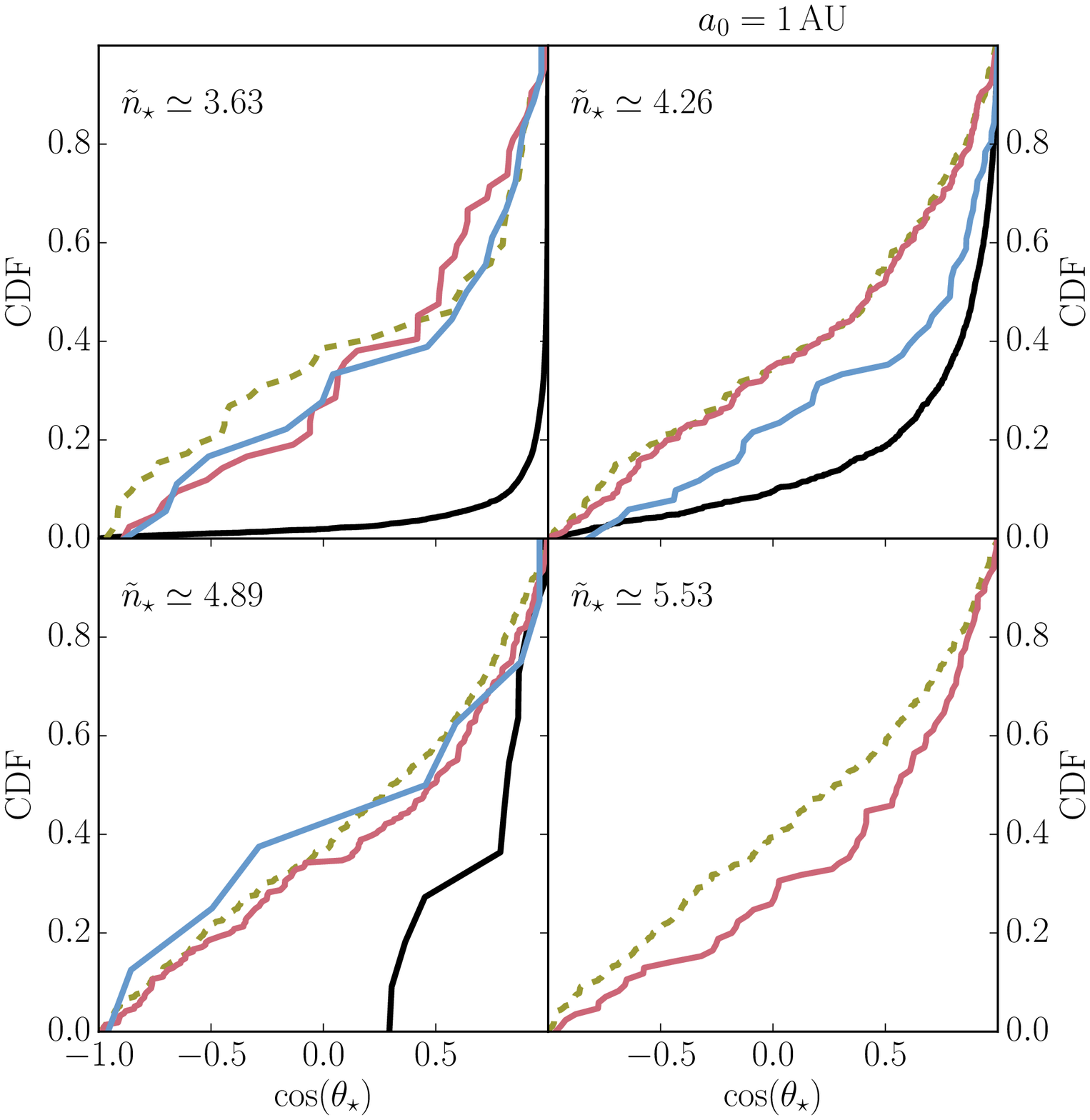}
}
\caption {Obliquity distributions for the HJ, WJ, tidal disruption, and no migration outcomes in the simulations and for four densities, combining data from the different $\renc$. The `no migration' and WJ curves are not plotted at the highest density, because these outcomes were too rare to produce good statistics. The top (bottom) panels show the probability (cumulative) distributions. }
\label{fig:stellar_obliquities}
\end{figure}

The stellar obliquity $\theta_\star$, the angle between the stellar spin and planetary orbit, is known to be broadly distributed for field HJs, with almost half of HJs having obliquities that exceed 10--20$^\circ$ and a handful having obliquities as high as $\sim 180^\circ$ (see \citealt{2015ARA&A..53..409W} for a review). In \F\,\ref{fig:stellar_obliquities}, we show the obliquity distributions for the HJ, WJ, tidal disruption, and no migration outcomes in the simulations with $a_0=1\,\au$, combining data from the different $\renc$. The top (bottom) panels show the probability density (cumulative) distributions. In the simulations, the spin of the star was not modeled; to compute the obliquity, we assume that the stellar spin and planetary orbit were initially aligned, and that the stellar spin direction is fixed. 

For all of the four outcomes, the final obliquity distribution is quite broad and not far from isotropic (i.e., a flat distribution in  $\cos \theta_\star$). For nonmigrating planets, the final distributions are not exactly flat, i.e., some memory of the initial zero obliquity is retained. Stronger interactions are involved in the other outcomes (in particular, HJ and WJ formation); consequently, the obliquity distributions for these cases tend to be more isotropic. The fraction of retrograde HJs ranges from $\approx 0.2$ to $\approx 0.4$, higher than the few per cent observed in the field \citep{2015ARA&A..53..409W}. A high fraction of retrograde obliquities would be a telltale sign for an encounter-induced high-$e$ migration origin of HJs in dense clusters.

\subsubsection{Stopping times}
\label{sect:pop_syn:times}

\begin{figure}
\center
\iftoggle{ApJFigs}{
\includegraphics[scale = 0.45, trim = 10mm 0mm 0mm 0mm]{end_times_run05.eps}
}{
\includegraphics[scale = 0.45, trim = 10mm 0mm 0mm 0mm]{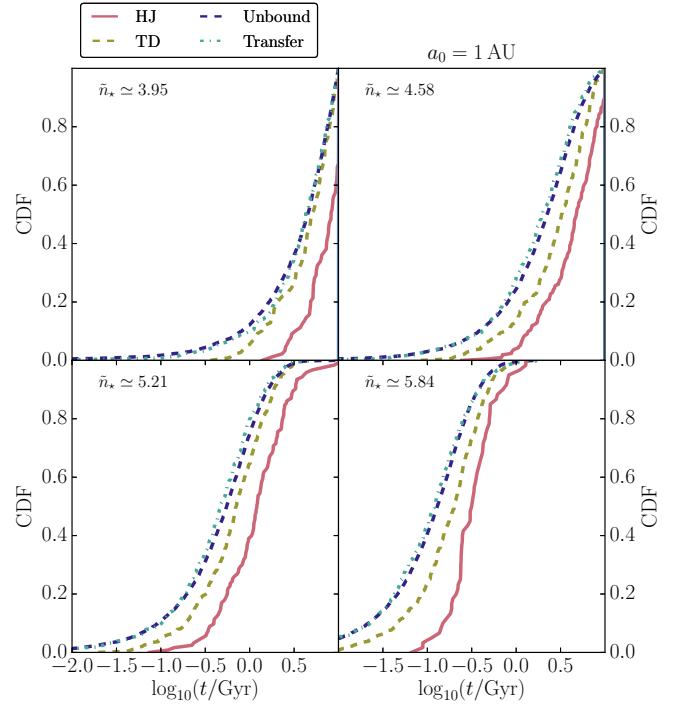}
}
\caption { Cumulative distributions of the stopping times, i.e., the times required to form an HJ, to tidally disrupt the planet, to unbind the planet, or to transfer the planet, combining different data for the different $\renc$, and for $a_0=1\,\au$. }
\label{fig:formation_times}
\end{figure}

The stopping times in the simulations are the times needed for encounters to form an HJ, to tidally disrupt the planet, to unbind the planet, or to transfer the planet. In \F\,\ref{fig:formation_times}, we show the cumulative distributions of these times for different densities, combining data for the different $\renc$ (with $a_0=1\,\au$). As expected, the stopping times are strongly dependent on the density: a higher density implies that the aforementioned stopping conditions occur earlier. Generally, the unbound and transfer outcomes occur earlier than the tidal disruption outcomes, which in turn occur earlier than the HJ outcomes. This can be understood by noting that unbinding or transfer generally occurs through a single strong encounter, whereas tidal disruption may involve a larger number of weaker encounters. Finally, the formation of HJs tends to occur later than tidal disruptions, given that some tidal evolution is typically involved to produce the HJ, adding to the formation time. 

The stopping times are generally shorter for larger $a_0$ (results not shown here), which can be understood qualitatively from the stronger effect of perturbers if the planet is less bound to its host star.

\subsubsection{Transferred planets}
\label{sect:pop_syn:trans}

\begin{figure}
\center
\iftoggle{ApJFigs}{
\includegraphics[scale = 0.45, trim = 0mm 0mm 0mm 0mm]{transferred_planets_run05.eps}
}{
\includegraphics[scale = 0.45, trim = 0mm 0mm 0mm 0mm]{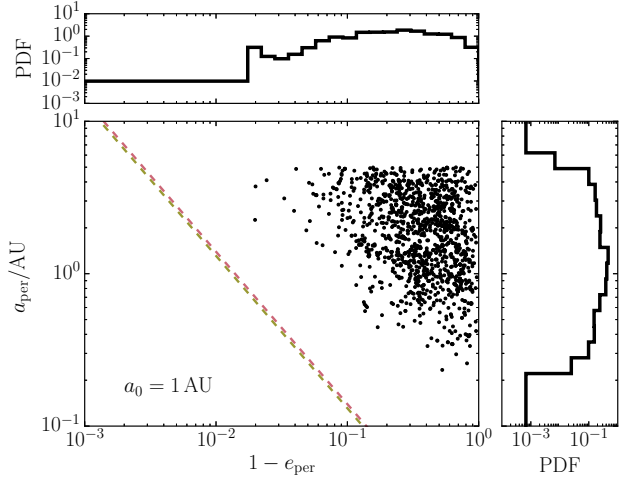}
}
\caption { Semimajor axis $a_\mathrm{per}$ and eccentricity $e_\mathrm{per}$ of the planet with respect to the perturber to which it was transferred, combining results from all densities and encounter sphere radii and assuming $a_0=1\,\au$.  To reduce clutter, only a fifth of the available data points are shown. The top and right panels show the marginalized eccentricity and semimajor axis distributions, respectively. The yellow line shows the assumed tidal disruption limit, and the red dashed line shows $r_\mathrm{p} = 3 \,\rsun$, approximately the periapsis distance for which tidal effects are important. }
\label{fig:transferred_planets}
\end{figure}

The planet is transferred to a perturbing star in up to $\approx 5\%$ of our simulations with $a_0=1\,\au$ (see \F\,\ref{fig:fractions}). In \F\,\ref{fig:transferred_planets}, we show the semimajor axis $a_\mathrm{per}$ and  eccentricity $e_\mathrm{per}$ of the planet with respect to the perturber to which it was transferred, combining results from all densities and encounter sphere radii, and assuming $a_0=1\,\au$. The top and right panels show the marginalized eccentricity and semimajor axis distributions, respectively. The yellow line shows the assumed tidal disruption limit, and the red dashed line shows $r_\mathrm{p} = 3 \,\rsun$, approximately the periapsis distance for which tidal effects are important. The captured planets have semimajor axes ranging from roughly 0.2--5 $\au$ (the upper limit is an artifact of the definition described in \S\,\ref{sect:pop_syn:sc}). The bulk of the transferred planets are not captured onto orbits for which tidal evolution is immediately important (of course, subsequent perturbations by encounters with other stars could drive tidal migration at later stages).

\subsubsection{Perturber properties}
\label{sect:pop_syn:per}
\begin{figure}
\center
\iftoggle{ApJFigs}{
\includegraphics[scale = 0.45, trim = 10mm 0mm 0mm 0mm]{perturber_Qs_CDF_run05.eps}
}{
\includegraphics[scale = 0.45, trim = 10mm 0mm 0mm 0mm]{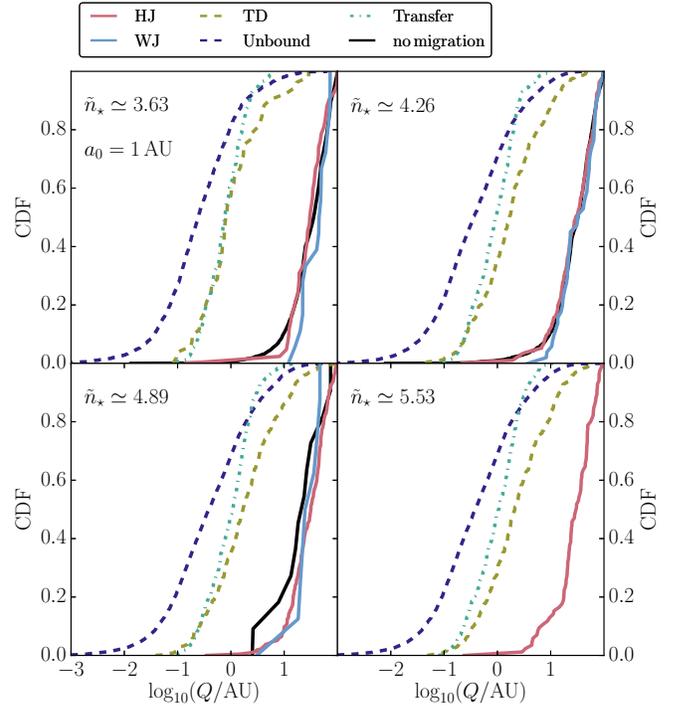}
}
\caption { Cumulative distributions of $Q$ for the last perturber associated with each outcome. Data are combined for the different $\renc$, and $a_0=1\,\au$. Note that there are no WJ or `no migration' outcomes for the highest density shown. }
\label{fig:perturber_Qs}
\end{figure}

\begin{figure}
\center
\iftoggle{ApJFigs}{
\includegraphics[scale = 0.45, trim = 10mm 0mm 0mm 0mm]{perturber_Ms_CDF_run05.eps}
}{
\includegraphics[scale = 0.45, trim = 10mm 0mm 0mm 0mm]{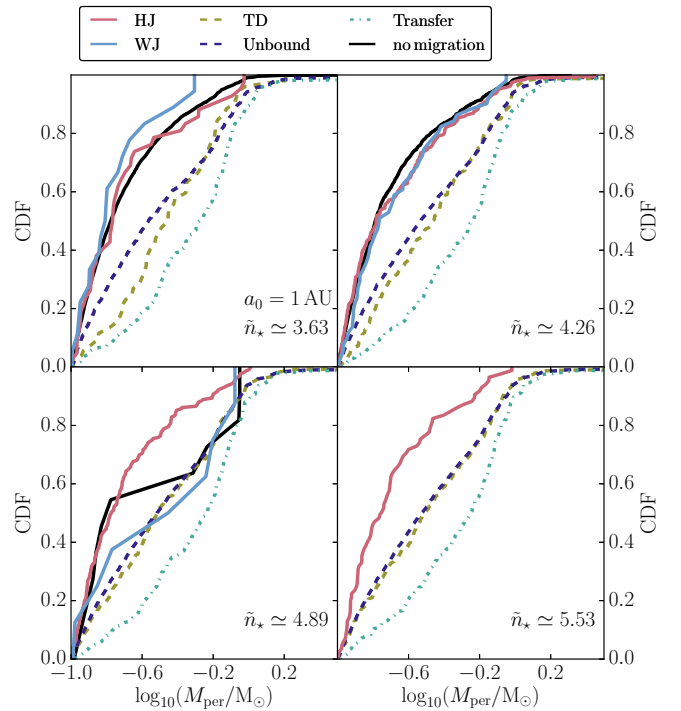}
}
\caption { Similar to \F\,\ref{fig:perturber_Qs}, now showing the cumulative distributions of the perturber masses, $\mper$. }
\label{fig:perturber_Ms}
\end{figure}

\begin{figure}
\center
\iftoggle{ApJFigs}{
\includegraphics[scale = 0.45, trim = 10mm 0mm 0mm 0mm]{perturber_V_inftys_CDF_run05.eps}
}{
\includegraphics[scale = 0.45, trim = 10mm 0mm 0mm 0mm]{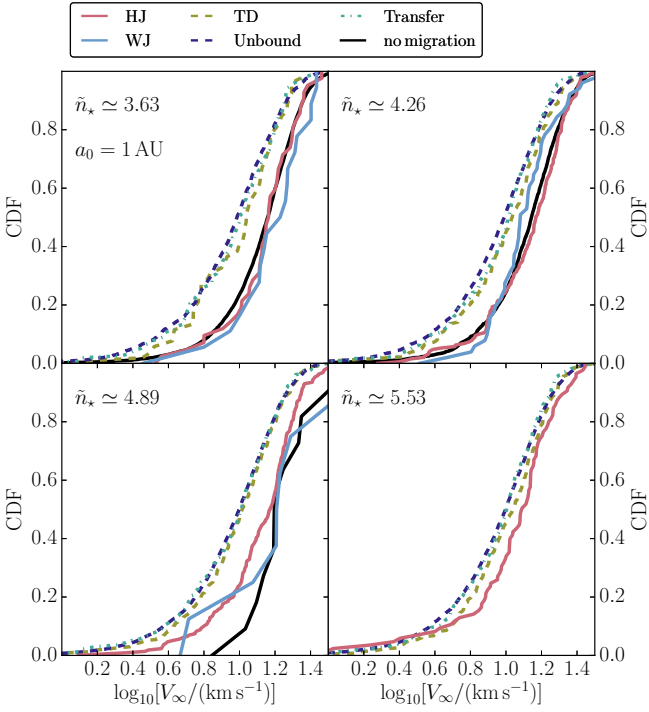}
}
\caption { Similar to \F\,\ref{fig:perturber_Qs}, now showing the cumulative distributions of the perturber speeds at infinity, $V_\infty$.}
\label{fig:perturber_V_inftys}
\end{figure}

In Figs.\,\ref{fig:perturber_Qs}, \ref{fig:perturber_Ms} and \ref{fig:perturber_V_inftys}, we illustrate how the simulation outcomes are determined by the perturber properties. In particular, in these three figures we show the cumulative distributions of $Q$, $\mper$ and $V_\infty$, respectively, for the last perturber associated with each outcome. For example, in the case of the unbound outcome, $Q$, $\mper$, and $V_\infty$ correspond to the perturber that triggered the ejection of the planet. As may be expected intuitively, the `strong' outcomes (tidal disruption, unbinding or transfer) tend to be associated with lower values of $Q$, higher values of $\mper$, and lower values of $V_\infty$. Note that for the HJs, the last encounter is usually not the encounter that drove tidal migration: typically, one or several strong encounters drive high eccentricity and tidal migration, and subsequently the planet is decoupled from further perturbations. The figures show the properties of the last perturber prior to the HJ stopping condition. The figures shown here apply only to the case $a_0=1\,\au$, but the perturber properties are not much changed in the simulations with larger $a_0$.

\subsection{Analytic model for encounter-induced high-$e$ migration}
\label{sect:pop_syn:an}

\begin{figure}
\center
\iftoggle{ApJFigs}{
\includegraphics[scale = 0.46, trim = 0mm -5mm 0mm 0mm]{a_e_plane_run05_ex.eps}
}{
\includegraphics[scale = 0.46, trim = 0mm -5mm 0mm 0mm]{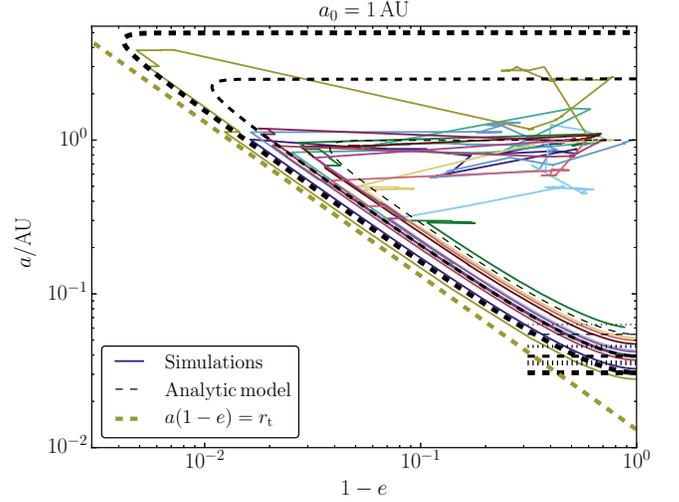}
}
\caption { Colored solid lines: tracks in the $(a,1-e)$ diagram of several individual systems from the simulations that lead to HJs. For all tracks, $a_0=1\au$, the density is $n_\star \simeq 2.8\times10^4 \, \mathrm{pc^{-3}}$, and $\renc$ is either 25, 50, 75 or 100 $\au$. The black dashed lines show analytic results from \S\,\ref{sect:pop_syn:an} (eq.~\ref{eq:diff_a_ell_no_time}; setting $\mper = 0.26 \, \mathrm{M}_\odot$, approximately the mean perturber mass in the simulations), for three different values of the initial semimajor axis: 1, 2.5 and 5 $\au$ (line thickness increases with increasing initial semimajor axis). In the lower-right part of the figure, the short black horizontal dotted lines show the simplest estimate of the stalling semimajor axis, equation~(\ref{eq:a_p_f_rough}); the short black horizontal dashed lines show the more accurate estimate, equation~(\ref{eq:a_p_f_better}). The yellow dashed line is the tidal disruption line, $a(1-e)=r_\mathrm{t}$ with $r_\mathrm{t}$ given by equation~(\ref{eq:r_t}). }
\label{fig:a_e_plane}
\end{figure}

In this section, we discuss a simplified analytic model for the evolution of the planetary orbit including perturbations by encounters and tidal evolution (assuming equilibrium tides, as usual). The main aim is to obtain an estimate of the final `stalling' semimajor axis of the HJs, which was observed in the simulations to be $\approx 0.03\,\au$ (\S\,\ref{sect:pop_syn:orb} and \F\,\ref{fig:HJ_periods_obs}). 

First, we show in \F\,\ref{fig:a_e_plane} with colored solid lines tracks in the $(a,1-e)$ plane of several individual systems from the simulations with $a_0=1\,\au$. All systems start at some location on the right part of the horizontal line at $a = 1 \au$ (the initial eccentricity distribution was assumed to be a Rayleigh distribution; see Table\,\ref{table:par}). Due to encounters, the semimajor axis and eccentricity change randomly, and the planet random-walks in the $(a,1-e)$ plane. For the examples shown, eventually the periapsis distance becomes sufficiently small for tidal evolution to become important, after which the system evolves with approximately constant angular momentum or semilatus rectum $a\left(1-e^2\right)$ until the orbit is circularized, at $a = a_\mathrm{f}$.

To model this type of evolution, we consider the coupled differential equations for $a$ and $\ell$, where $\ell \equiv \sqrt{1-e^2}$ is the normalized angular momentum. We approximate the orbit-averaged tidal evolution equations~(\ref{eq:TF_av}) by replacing $f(e)$ by $f(1)$ (since we are mainly interested in the competition between tidal circularization and eccentricity excitation by encounters at high eccentricity), giving
\begin{align}
\label{eq:dif_a_ell_TF}
\left \{
\begin{array}{l>{\displaystyle}l}
\dot{a}_{\mathrm{TF}} &\simeq - \beta_a \frac{a}{\tau_\mathrm{TF}} \left ( \frac{R_\mathrm{p}}{a} \right )^8 \ell^{-15} \left(1-\ell^2\right); \\
\dot{\ell}_{\mathrm{TF}} &\simeq \beta_e \frac{1}{\tau_\mathrm{TF}} \left ( \frac{R_\mathrm{p}}{a} \right )^8 \ell^{-14} \left(1-\ell^2\right), \\
\end{array}
\right .
\end{align}
where the overdots denote derivatives with respect to time. The (constant) tidal friction timescale $\tau_\mathrm{TF}$ is defined in terms of the tidal time lag $\tau_\mathrm{p}$ according to
\begin{align}
\tau_\mathrm{TF}^{-1} \equiv \frac{ 21 \, k_\mathrm{AM,p} G M_\star \tau_\mathrm{p}}{R_\mathrm{p}^3} \frac{M_\star}{M_\mathrm{p}};
\end{align}
the (constant) factor $\beta_a$ is given by 
\begin{align}
\beta_a \equiv f(1) = \frac{4059}{1120},
\end{align}
and $\beta_e = \beta_a/2$.

To estimate the orbital changes due to encounters, we make the simplifying assumption that the encounters are purely secular and thus do not exchange energy, i.e., we ignore changes of the semimajor axis and only consider eccentricity changes. Evidently, energy changes are important as shown by the random walks in semimajor axis in \F\,\ref{fig:a_e_plane}. However, we note that strong encounters that significantly change the semimajor axis tend to tidally disrupt the planet, unbind the planet, or transfer the planet to the perturber. In other words, the encounter history of an HJ generally does not include strong encounters associated with large energy changes, otherwise the planet would likely not survive and remain bound to the host star.

In the secular approximation, the rate of change of eccentricity due to encounters, $\dot{e}_\mathrm{enc}$, can be estimated in the quadrupole-order approximation according to (Hamers, in preparation)
\begin{align}
\dot{e}_\mathrm{enc} &\simeq  \frac{15}{2} n \frac{\mper}{M_\star} \left ( \frac{a}{R} \right )^3 \frac {\left ( \ve{e} \cdot \ve{R} \right ) \left [ \unit{e} \cdot \left (\ve{\jmath} \times \ve{R} \right ) \right ]}{R^2}.
\end{align}
Here, $\ve{e}$ and $\ve{\jmath}\equiv \sqrt{1-e^2} \, \unit{\jmath}$ are the eccentricity and the dimensionless angular-momentum vectors, respectively, and we assumed $M_\mathrm{p} \ll M_\star$ (otherwise, $M_\star$ should be replaced by $M_\star + M_\mathrm{p}$). Averaging over all orientations of the encounter (with $R$ fixed) gives $\langle \dot{e}_\mathrm{enc} \rangle = 0$, as expected: on average, encounters do not induce a net change in the scalar eccentricity. To estimate the typical value of $\dot{e}_\mathrm{enc}$, we compute the rms value averaged over all orientations, i.e.,
\begin{align}
\label{eq:dot_ep_step1}
\left \langle \dot{e}^2_\mathrm{enc} \right \rangle^{1/2} \simeq \frac{\sqrt{15}}{2} e \sqrt{1-e^2} \, n \left ( \frac{a}{R} \right )^3 \frac{\mper}{M_\star}.
\end{align}
Equation~(\ref{eq:dot_ep_step1}) estimates the typical eccentricity change per unit time during an encounter when the separation is $R$. To obtain the time-averaged rate of eccentricity change, we must multiply equation~(\ref{eq:dot_ep_step1}) by a `duty cycle' factor, defined as
\begin{align}
\label{eq:f_duty_def}
f_\mathrm{duty} \equiv \frac{\Delta t_\mathrm{passage}(R)}{\Delta t_\mathrm{enc}(R)},
\end{align}
where $\Delta t_\mathrm{passage}(R)$ is the duration of an encounter and $\Delta t_\mathrm{enc}(R)$ is the timescale for the next encounter to occur. Approximately, $\Delta t_\mathrm{passage}(R)\simeq R/\srel$, and $\Delta t_\mathrm{enc} (R)\sim 1/\Gamma(R)$, with $\Gamma(R)$ estimated with equation~(\ref{eq:Gamma_limit}) after $\renc$ is replaced by $R$. These estimates for the timescales in equation~(\ref{eq:f_duty_def}) imply
\begin{align}
\label{eq:f_duty}
f_\mathrm{duty} \simeq 2\sqrt{2\pi} \, R^3 n_\star,
\end{align}
independent of $\srel$ (a higher velocity dispersion implies a shorter encounter timescale  $\Delta t_\mathrm{enc}$, but this is compensated equally by a shorter passage timescale $\Delta t_\mathrm{passage}$). Note that $f_\mathrm{duty}$ can also be obtained from equation~(\ref{eq:ratio_t_pas_t_enc}) by replacing $\renc$ by $R$. Writing equation~(\ref{eq:dot_ep_step1}) in terms of $\ell$, and including $f_\mathrm{duty}$ give
\begin{align}
\label{eq:ell_p_dot_enc_est}
\nonumber \left \langle \dot{\ell}^2_\mathrm{enc}  \right \rangle^{1/2}_\mathrm{eff} &\simeq \frac{\sqrt{15}}{2} \, n \frac{\mper}{M_\star} \left ( \frac{a}{R} \right )^3 \left(1-\ell^2\right ) f_\mathrm{duty}\\
&\simeq \sqrt{30\pi} \left ( G M_\star a^3 \right )^{1/2} \frac{\mper}{M_\star} n_\star \left(1-\ell^2\right ).
\end{align}
Note that the result is independent of $R$; this scaling arises because the rate of encounters with closest approach less than $R$ is $\propto R^2$, the torque from an encounter is $\propto 1/R^3$, and the duration is $\propto R$.

Adding equation~(\ref{eq:ell_p_dot_enc_est}) to equation~(\ref{eq:dif_a_ell_TF}), with a minus sign to model the cases where the eccentricity random-walks to higher values, we get
\begin{align}
\label{eq:dif_a_ell}
\left \{
\begin{array}{l>{\displaystyle}l}
\dot{a} &\simeq - \beta_a \frac{a}{\tau_\mathrm{TF}} \left ( \frac{R_\mathrm{p}}{a} \right )^8 \ell^{-15} \left(1-\ell^2 \right ); \\
\dot{\ell} &\simeq \beta_e \frac{1}{\tau_\mathrm{TF}} \left ( \frac{R_\mathrm{p}}{a} \right )^8 \ell^{-14} \left(1-\ell^2 \right ) \\
&\quad- \sqrt{30\pi} \left ( G M_\star a^3 \right )^{1/2} \frac{\mper}{M_\star} n_\star \left(1-\ell^2 \right ). \\
\end{array}
\right .
\end{align}
Division of the two expressions gives
\begin{align}
\label{eq:diff_a_ell_no_time}
\frac{\mathrm{d} \ell}{\mathrm{d} (a/a_{0})} \simeq - \frac{1}{2} \frac{\ell}{a/a_{0}} \left [ 1 - A \left (\frac{a}{a_{0}} \right )^\frac{19}{2} \ell^{14} \right ],
\end{align}
where $a_{0}$ is a constant, and the dimensionless constant $A$ is defined as
\begin{align}
A \equiv \sqrt{30\pi} \left ( \frac{G M_\star}{a_{0}^3} \right )^\frac{1}{2} \frac{\mper}{M_\star} \left ( \frac{a_{0}}{R_\mathrm{p}} \right )^8 a_{0}^3 n_\star \frac{\tau_\mathrm{TF}}{\beta_e}.
\end{align}
For the purposes below, it is convenient to define $a_{0}$ as the initial semimajor axis and to assume that the initial eccentricity is zero, such that the initial condition in the differential equation~(\ref{eq:diff_a_ell_no_time}) is $(a/a_0,\ell)_\mathrm{init} = (1,1)$. 

Equation~(\ref{eq:diff_a_ell_no_time}) can be solved analytically. Before doing so, we remark that an approximate stalling semimajor axis can be obtained from equation~(\ref{eq:diff_a_ell_no_time}) without solving explicitly for $\ell(a)$. To achieve this, note that $\dot{\ell} = 0$ at a critical value of $\ell$, which we call $\ell_{\mathrm{c}}$. In evaluating this critical value, however, it is a good approximation to set $a = a_{0}$, i.e., the semimajor axis typically does not decrease much before the orbit reaches the critical $\ell$, and $1-\ell^2 \approx 1$ (see Fig.~\ref{fig:a_e_plane}). Requiring $\dot{\ell}=0$ in equation~(\ref{eq:dif_a_ell}) with $a = a_{0}$ and assuming $1-\ell^2 = 1$ gives
\begin{align}
\ell_{\mathrm{c}} \simeq A^{-1/14}.
\end{align}
Assuming that after reaching $\ell = \ell_{\mathrm{c}}$ the evolution is completely dominated by tides, the final semimajor axis can be approximated using the constancy of the semilatus rectum, giving
\begin{align}
\label{eq:a_p_f_rough}
a_{\mathrm{f}} \simeq a_{0} \ell_{\mathrm{c}}^2 \simeq a_{0} A^{-1/7}.
\end{align}
In the bottom-right part  of Fig.~\ref{fig:a_e_plane}, the estimate (\ref{eq:a_p_f_rough}) is shown with the short black horizontal dotted lines (setting $\mper = 0.26 \, \mathrm{M}_\odot$, approximately the mean perturber mass in the simulations). Note that semimajor axis changes due to encounters are not included in our model, whereas in some cases in the simulations the semimajor axis has changed before tidal evolution becomes important. Therefore, we show results of the analytic model for three values of the initial semimajor axis, i.e., either 1, 2.5, or 5 $\au$. 

Next, we obtain a slightly more accurate estimate by explicitly solving equation~(\ref{eq:diff_a_ell_no_time}) for $\ell$, as a function of $a$. The solution is
\begin{align}
\label{eq:ell_p_sol}
\ell(a) \simeq 5^{1/14} \left ( \frac{a}{a_{0}} \right )^{-\frac{1}{2}} \left [ 5 - 14 A \left \{ \left ( \frac{a}{a_{0}} \right )^{\frac{5}{2}} - 1 \right \} \right ]^{-\frac{1}{14}}.
\end{align}
To obtain the stalling semimajor axis, $a_{\mathrm{f}}$, we set $\ell(a_{\mathrm{f}}) = 1$ and assume $a_{\mathrm{f}} \ll a_{0}$, giving
\begin{align}
\label{eq:a_p_f_better}
\nonumber a_{\mathrm{f}} &\simeq a_{0} \, 5^{1/7} \left ( 5 + 14 A \right )^{-1/7} \simeq a_{0} \, (5/14)^{1/7} A^{-1/7} \\
\nonumber &= a_{0} \left ( \frac{5}{14} \right )^{\frac{1}{7}} \biggl [ \frac{1}{\sqrt{30\pi}} \frac{4059}{2240} \frac{21 \, k_\mathrm{AM,p} G M_\star \tau_\mathrm{p}}{R_\mathrm{p}^3} \left ( \frac{M_\star}{M_\mathrm{p}} \right ) \\
&\quad  \times \left ( \frac{M_\star}{\mper} \right ) \left ( \frac{a_{0}^3}{G M_\star} \right )^{\frac{1}{2}} \left ( \frac{R_\mathrm{p}}{a_{0}} \right )^8 \left ( a_{0}^{3} n_\star \right )^{-1} \biggl ]^{\frac{1}{7}}.
\end{align}
We assumed $A\gg 1$, which is the case here (in Fig.~\ref{fig:a_e_plane}, $A \simeq 2.5 \times 10^8$, $1.5\times 10^{12}$ and $1.1\times 10^{15}$ for the three adopted values of the initial semimajor axis). Equation~(\ref{eq:a_p_f_better}) is the same as equation~(\ref{eq:a_p_f_rough}) apart from a factor of $(5/14)^{1/7} \simeq 0.863$. The more refined estimate for the stalling semimajor axis $a_\mathrm{f}$, equation~(\ref{eq:a_p_f_better}), is shown with the short black horizontal dashed lines. The analytic solutions (\ref{eq:ell_p_sol}) are shown with the black dashed lines. 

Although a large number of assumptions and approximations were made in its derivation, equation~(\ref{eq:ell_p_sol}) gives a reasonable representation of the typical evolution in the $(a,1-e)$ plane, and of the stalling semimajor axis.  Our estimate for $a_\mathrm{f}$ is insensitive to the poorly known tidal time lag $\tau_\mathrm{p}$ since it only appears with a small exponent ($\frac{1}{7}$). Also, $a_\mathrm{f}$ depends only weakly on the initial semimajor axis $a_0$ and stellar number density $n_\star$: $a_\mathrm{f}\propto a_0^{-5/14}$ and $a_\mathrm{f} \propto n_\star^{-1/7}$. 

The stalling semimajor axes and the corresponding orbital periods according to equation~(\ref{eq:a_p_f_better}) were shown with the black vertical dashed lines in Figures \ref{fig:HJ_smas} and \ref{fig:HJ_periods_obs}, respectively, where $a_0$ was set to the initial semimajor axis in the simulations. These lines are comparable to the results from the simulations and follow the trends with $n_\star$ and $a_0$, although the simulations tend to produce HJs with shorter orbital periods. This can be understood in part by noting that  the semimajor axis tends to increase in the simulations before tidal evolution takes over and larger initial semimajor axes in the model result in smaller stalling distances. 

In equation~(\ref{eq:a_p_f_better}), the strongest dependence of the stalling semimajor axis is on the planetary radius, $a_{\mathrm{f}}\propto R_\mathrm{p}^{5/7}$. A similar dependence is seen in HJs formed by high-$e$ migration in stellar binaries or multiplanet systems \citep{2011ApJ...735..109W,2015ApJ...799...27P,2015ApJ...805...75P}. This property can be attributed to the strong dependence of the efficiency of tidal circularization on $R_\mathrm{p}$.

Comparing high-$e$ migration in stellar binaries to our scenario, we note that in the former case, the maximum eccentricity (and hence the stalling radius) generally arises from the quenching of secular oscillations by precession due to short-range forces (e.g., general relativity). In our scenario, the stalling radius is set by a competition between periapsis decrease due to encounters and circularization due to tides. If short-range forces were absent in stellar binaries or multiplanet systems, then the maximum eccentricity would similarly be limited by the competition between eccentricity driving and tidal circularization.

\section{Discussion}
\label{sect:discussion}

\subsection{Migration fractions}
\label{sect:discussion:migr_fr}
The migration fractions in our simulations (the sum of $f_\mathrm{HJ}$, $f_\mathrm{WJ}$ and $f_\mathrm{TD}$) increase monotonically with stellar density (Fig.~\ref{fig:fractions}). For initial semimajor axis $a_0=1\,\au$, the migration fractions are up to a few per cent for densities of $n_\star \sim 10^4 \, \mathrm{pc^{-3}}$, and increase to $\approx 6\%$ for $n_\star=10^6\,\mathrm{pc^{-3}}$. The HJ fractions are below $\approx 1\%$ for $n_\star \lesssim 10^4\,\mathrm{pc^{-3}}$,  increase to $\approx 2\%$ for $n_\star \approx 4\times 10^4\,\mathrm{pc^{-3}}$, and decrease again to below $\approx 1\%$ at $n_\star=10^6\,\mathrm{pc^{-3}}$. The WJ fractions are nearly zero for $n_\star \lesssim 2 \times 10^3\,\mathrm{pc^{-3}}$ and $n_\star \gtrsim 10^5\,\mathrm{pc^{-3}}$, and reach a maximum of $\approx 0.5\%$ at $n_\star \approx 2\times 10^4\,\mathrm{pc^{-3}}$. Generally, the migration fractions are lower for larger $a_0$. For $a_0=2 \, \au$, the highest HJ fraction is $\approx 1.5\%$ at $n_\star \approx 2\times 10^4\,\mathrm{pc^{-3}}$, whereas for $a_0=4\,\au$ the highest HJ fraction is $\approx 1\%$ at $n_\star \approx 1\times 10^4\,\mathrm{pc^{-3}}$.

Adopting an HJ fraction of 0.02 from the simulations with $n_\star \approx 4\times 10^4\,\mathrm{pc^{-3}}$ and assuming a conservative giant-planet fraction of 0.05 \citep{2008PASP..120..531C}, our simulations predict that $\sim 0.1\%$ of solar-type stars in the central regions of some GCs could host HJs that formed through this mechanism. This fraction is an order of magnitude lower than the observed HJ fraction around solar-type field stars, $\sim 1\%$ (e.g., \citealt{2012ApJ...753..160W}). We note that higher HJ fractions would be obtained if (i) the fraction of GC stars that form giant planets is larger than 0.05; (ii) the efficiency of tidal dissipation in the giant planet, which is still uncertain, is larger than what we have assumed; and (iii) the planetary radius was inflated, as is observed for HJs around field stars (e.g., \citealt{2011ApJ...729L...7L}). 

Our simulated HJ fractions are similar to those found by \citet{2016ApJ...816...59S} for open clusters. Thus, the apparently more hostile environment of GCs (typically higher densities and longer cluster ages) does not imply that the efficiency of HJ production through high-$e$ migration is lower in GCs compared to open clusters.

\subsection{Obliquities}
\label{sect:discussion:obl}

The obliquity distribution of HJs is a powerful probe of their origin. As shown in \S\,\ref{sect:pop_syn:obl}, the obliquity distribution of HJs formed through hyperbolic encounters tends to be isotropized. In the case of high-$e$ migration in stellar binaries, two peaks in the obliquity distribution are expected near $\sim 40^\circ$ and $\sim 130^\circ$ (\citealt{2007ApJ...669.1298F,2014ApJ...793..137N,2016MNRAS.456.3671A}; see \citealt{2017MNRAS.465.3927S} for a detailed study of the origin of the bimodal distribution). For high-$e$ migration in stellar triples, the obliquity distribution also shows two broad peaks near these values \citep{2017MNRAS.466.4107H}. For secular chaos in multiplanet systems, the obliquity distribution has a width of a few tens of degrees \citep{2014PNAS..11112610L,2017MNRAS.464..688H}. The obliquity distribution could therefore distinguish HJ formation by disk migration and various high-$e$ migration mechanisms, but of course this requires, first, the detection of a large sample of  HJs in GCs and, second, the measurement of their obliquities --- both challenging tasks.

\subsection{Other effects related to encounters}
\label{sect:discussion:other}

We have not considered all of the possible perturbations of planetary systems in dense stellar systems, such as the centers of GCs. Binary stars are known to exist in GCs (either formed primordially or dynamically), and encounters with binary stars can be much more effective than encounters with single stars (e.g., \citealt{2015MNRAS.448..344L}). Also, the dynamics are richer in the case of multiplanet systems. For example, \citet{2004AJ....128..869Z} showed that perturbations from stellar encounters can propagate inwards from the outermost planets, inducing high eccentricities in the innermost planet even if the stellar encounters are not effective at exciting the innermost planet directly. Similar effects applied to star clusters were considered by \citet{doi:10.1093/mnras/stx1464}, who found that planet-planet interactions enhance the rate of planet ejection.

\subsection{Explanation for the HJ window}
\label{sect:discussion:HJ_window}
As discussed previously, HJs are produced in our simulations only for a limited range of densities. Evidently, if the density is low, the probability for strong encounters to affect either the semimajor axis or the eccentricity is low as well, and no HJs are produced. For high densities, the reason why the HJ fraction decreases is less clear. Here, we provide an explanation. Typically, once the periapsis distance has been reduced by encounters to a few stellar radii (but still larger than the tidal disruption distance $r_\mathrm{t}$), tidal evolution starts to reduce the semimajor axis and circularize the orbit, eventually producing an HJ on some time scale, $\Delta t_\mathrm{tide}$ (see the top-left panel of \F\,\ref{fig:examples} for an example). However, if the stellar density is high, it is possible that a strong encounter occurs before the orbit is completely circularized, leading to the ejection or tidal disruption of the planet.

The above argument can be quantified approximately by requiring that $\Delta t_\mathrm{tide}$ is equal to the time between encounters, $\Delta t_\mathrm{enc}$. This condition corresponds to a critical density, $n_{\star,\mathrm{crit}}$; if $n_\star > n_{\star,\mathrm{crit}}$, we expect that the efficiency of HJ formation is reduced by the above process. Therefore, the maximum in the HJ fraction should occur near $n_{\star,\mathrm{crit}}$. 

The tidal circularization time scale can be estimated from equations~(\ref{eq:dif_a_ell_TF}), which apply in the high-eccentricity limit. Using the constancy of the semilatus rectum, $a\left(1-e^2\right)$, the equation for $\dot{a}$ can be written as
\begin{align}
\label{eq:HJ_window_delta_t_tides}
\dot{a} \simeq - \frac{\beta_a}{128 \sqrt{2}\,\tau_\mathrm{TF}} \sqrt{a r_\mathrm{p,0}} \left ( \frac{R_\mathrm{p}}{r_\mathrm{p,0}} \right )^8,
\end{align}
where $r_\mathrm{p,0}$ is the initial periapsis distance, which we take to be $r_\mathrm{p,0} = 3 \, \rsun \simeq 0.014 \,\au$. Here, we assumed $r_\mathrm{p,0} \ll a_0$, where $a_0$ is the initial semimajor axis, which we set to $a_0=1\,\au$. Integrating equation~(\ref{eq:HJ_window_delta_t_tides}), we find that $a\to 0$ after a time 
\begin{align}
\Delta t_\mathrm{tide} \sim 256 \sqrt{2} \, \frac{\tau_\mathrm{TF}}{\beta_a} \left ( \frac{a_0}{r_\mathrm{p,0}} \right )^{\frac{1}{2}} \left ( \frac{r_\mathrm{p,0}}{R_\mathrm{p}} \right )^8.
\end{align}

The encounter time scale can be estimated as $\Delta t_\mathrm{enc} \sim 1/\Gamma$, with $\Gamma$ given by equation~(\ref{eq:Gamma_limit}). The requirement $\Delta t_\mathrm{enc} \sim \Delta t_\mathrm{tide}$ then yields the critical density,
\begin{align}
\label{eq:HJ_window}
\nonumber n_{\star,\mathrm{crit}} &\sim \frac{21 \, \beta_a k_\mathrm{AM,p} }{1024 \sqrt{\pi}} \frac{G M_\star \tau_\mathrm{p}}{\renc^2 \srel R_\mathrm{p}^3} \frac{M_\star}{M_\mathrm{p}} \left ( \frac{r_\mathrm{p,0}}{a_0} \right )^{\frac{1}{2}} \left ( \frac{R_\mathrm{p}}{r_\mathrm{p,0}} \right )^8 \\
\nonumber &\simeq 3.7 \times 10^4 \, \mathrm{pc^{-3}} \, \left ( \frac{k_\mathrm{AM,p}}{0.25} \right )  \left ( \frac{\tau_\mathrm{p}}{0.66\,\mathrm{s}} \right ) \left ( \frac{M_\star}{1\,\msun} \right)^2 \\
\nonumber &\quad \times \left ( \frac{M_\mathrm{p}}{1 \, M_\mathrm{J}} \right )^{-1}  \left ( \frac{R_\mathrm{p}}{1\,R_\mathrm{J}} \right )^5 \left ( \frac{\renc}{50\,\au} \right )^{-2}  \\
&\quad \times \left ( \frac{\srel}{ \sqrt{2}\times 6 \,\mathrm{km \, s^{-1}} } \right )^{-1}  \left ( \frac{r_\mathrm{p,0}}{3 \, \rsun} \right )^{-\frac{15}{2}}  \left ( \frac{a_0}{1\,\au} \right )^{-\frac{1}{2}}.
\end{align}
Here, we substituted typical values from the simulations. In the simulations, the HJ fraction peaks at a density of a few times $10^4\,\mathrm{pc^{-3}}$, consistent with equation~(\ref{eq:HJ_window}). Furthermore, equation~(\ref{eq:HJ_window}) predicts that the peak should occur at lower densities for larger $a_0$. In particular, $n_{\star,\mathrm{crit}} \simeq 1.8 \times 10^4 \, \mathrm{pc^{-3}}$ for $a_0=4\au$ (with the other parameters unaltered). This is consistent with the bottom-right panel of \F\,\ref{fig:fractions}, which shows that the HJ fraction peaks around $1\times10^4\,\mathrm{pc^{-3}}$ if $a_0=4\,\au$.

\section{Conclusions}
\label{sect:conclusions}

We considered the formation of hot Jupiters (HJs) through high-eccentricity (high-$e$) migration in dense stellar systems, in particular the centers of globular clusters (GCs). A giant planet, which is assumed to form initially at a distance of a few $\au$ from its host star, is excited to high eccentricity and small periapsis distance by perturbations from passing stars, triggering migration driven by tides. Our main results are as follows:

\medskip \noindent 1. We presented a regularized restricted three-body code RR3 that can be used to calculate efficiently the effect of a passing perturber on a test particle (i.e., a planet) initially orbiting the host star. The equations of motion include tidal dissipation and general relativistic corrections. In a number of tests, we showed that our code gives a factor of $\sim 20$-100 performance increase with respect to a high-accuracy direct $N$-body code, without much loss of accuracy (\S\,\ref{sect:meth:val}).

\medskip \noindent 2. Using RR3, we carried out population-synthesis calculations of encounters with planetary systems in dense clusters. We found that the fraction of planets that evolve to become HJs can be as large as $\approx 2\%$ for number densities of $n_\star\approx4\times10^4\,\mathrm{pc^{-3}}$ and an initial planetary semimajor axis of $1\,\au$, and decreases to nearly zero for densities lower than  $10^3\,\mathrm{pc^{-3}}$ or higher than $10^6\,\mathrm{pc^{-3}}$. This window in stellar density can be explained qualitatively by noting that the density must be sufficiently high for encounters to decrease the periapsis distance within a Hubble time, but not so high that further encounters can eject the planet or drive it to tidal disruption before it migrates to smaller semimajor axes through tidal friction (see Section \ref{sect:discussion:HJ_window}). The dependence of the HJ fraction on the stellar density implies that HJs can form through high-$e$ migration in the centers of dense GCs, but not in their outskirts where the densities are lower. We emphasize that we have investigated a single channel for HJ formation in GCs; the true fraction of potentially observable transiting planets may be larger due to contributions from other migration channels, or from planets that may have formed in situ. 

\medskip \noindent 3. The HJ orbital period distribution in our simulations resembles the observed distribution around field stars in the solar neighborhood, but shifted to somewhat larger values:  the distribution in our simulations peak around 2.5 days, whereas the observed distribution peaks around 5 days. The similarity in periods reflects the strong dependence of tidal forces on distance, which requires that HJs are found close to their host star, independent of how their eccentricities were initially excited. 

\medskip \noindent 4. Warm Jupiters (WJs; giant planets with periods between 10 and 100 days) are produced in our simulations, although this process is relatively rare: the largest fraction is $\approx 0.5\%$, much smaller (relative to either nonmigrating planets or HJs) than observed in field solar-type stars \citep{2016A&A...587A..64S}. In contrast to secular formation scenarios in field stars, WJs can be formed in dense clusters in our simulations through encounters only, not requiring tidal interactions.

\medskip \noindent 5. The migration fraction (the sum of the HJ, WJ, and tidal disruption fractions) decreases with increasing initial semimajor axis of the planet. This can be understood qualitatively from the stronger effect of encounters if the planet is less tightly bound to its host star, implying an increased probability for the planet to become unbound. In particular, the highest HJ fraction (as a function of stellar density) decreases to $\approx 1.5\%$ for an initial semimajor axis of $2\,\au$, and to $\approx 1\%$ for an initial semimajor axis of $4\,\au$. 

\medskip \noindent 6. We presented a simplified analytic model for high-$e$ migration driven by encounters (\S\,\ref{sect:pop_syn:an}), which approximately captures the HJ period distributions in our simulations. In this model, excitation of the planetary eccentricity by secular encounters competes with tidal circularization and shrinkage.

\medskip \noindent 7. The obliquity distributions in our simulations tend toward an isotropic distribution. A fully isotropic distribution differs from the observed obliquity distributions for field HJs and high-$e$ migration models in, e.g., stellar binaries. The signature of a near-isotropic obliquity distribution may help to distinguish between high-$e$ migration models in GCs, if a number of HJs were found in these environments.

\medskip \noindent 8. In up to $\approx 5\%$ of our simulations, the planet is transferred to a perturbing star. The orbital elements of the transferred planet with respect to its new host star are widely distributed (see \F\,\ref{fig:transferred_planets}). Typically, the periapsis distance of the transferred planet is not small enough for significant tidal evolution. However, subsequent perturbations with other stars could drive tidal migration, producing captured HJs around stars with masses preferentially around $0.5\,\msun$ (see \F\,\ref{fig:perturber_Ms}). In some cases, planets can also be captured around more massive objects, i.e., compact objects such as neutron stars and black holes.

\section*{Acknowledgements}
We thank Re'em Sari, Yanqin Wu, Nathan Leigh, Kento Masuda, and Maxwell Cai for stimulating discussions, and the anonymous referee for a helpful report. A.S.H. gratefully acknowledges support from the Institute for Advanced Study, The Peter Svennilson Membership, and from NASA grant NNX14AM24G.

\bibliographystyle{mnras}
\bibliography{literature}

\end{document}